\documentclass{elsart}
\usepackage{amsmath}
\usepackage{graphicx}
\usepackage{subfigure}
\usepackage{algorithm}
\usepackage{algorithmic}
\usepackage{subfigure}
\usepackage{hyperref}

\newcounter{bla}
\newenvironment{refnummer}{%
\list{[\arabic{bla}]}%
{\usecounter{bla}%
 \setlength{\itemindent}{0pt}%
 \setlength{\topsep}{0pt}%
 \setlength{\itemsep}{0pt}%
 \setlength{\labelsep}{2pt}%
 \setlength{\listparindent}{0pt}%
 \settowidth{\labelwidth}{[9]}%
 \setlength{\leftmargin}{\labelwidth}%
 \addtolength{\leftmargin}{\labelsep}%
 \setlength{\rightmargin}{0pt}}}
 {\endlist}

\newcommand{\dtau}{\Delta\hspace{-.06cm}\tau}
\newcommand{\tr}{\operatorname{Tr}}
\newcommand{\re}{\operatorname{Re}}

\newcommand{\Qp}{Q^+}
\newcommand{\Qm}{Q^-}
\newcommand{\Qpm}{Q^\pm}
\newcommand{\hQp}{\hat Q^+}
\newcommand{\hQm}{\hat Q^-}
\newcommand{\hQpm}{\hat Q^\pm}

\begin{document}
\begin{frontmatter}

\title{tmLQCD: a program suite to simulate Wilson Twisted mass
  Lattice QCD}

\author[a]{Karl Jansen}\ and\ 
\author[b]{Carsten Urbach\thanksref{author}}

\thanks[author]{Corresponding author}

\address[a]{DESY, Patanenallee 6, Zeuthen, Germany}
\address[b]{Humboldt-Univertit{\"a}t zu Berlin, Institut f{\"u}r
  Physik,\\Newtonstr. 15, 12489 Berlin, Germany}

\begin{abstract}
  %Type your abstract here.
We discuss a program suite for simulating Quantum Chromodynamics 
on a 4-dimensional space-time lattice. The basic Hybrid Monte Carlo 
algorithm is introduced and a number of algorithmic improvements 
are explained. We then discuss the implementations of these concepts 
as well as our parallelisation strategy in the actual simulation code.
Finally, we provide a user guide to compile and run the program.
\begin{flushleft}
  %Insert your suggested PACS number here
\end{flushleft}

\begin{keyword}
Hybrid Monte Carlo algorithm; Lattice QCD;
  % Please give some freely chosen keywords that we can use in a
  % cumulative keyword index.
\end{keyword}

\begin{flushleft}
Preprint-Numbers: DESY 09-073, HU-EP-09/23, SFB/CPP-09-43
\end{flushleft}

\end{abstract}

\end{frontmatter}

% Computer program descriptions should contain the following
% PROGRAM SUMMARY.

{\bf PROGRAM SUMMARY}
  %Delete as appropriate.

\begin{small}
\noindent
{\em Manuscript Title: tmLQCD: a program suite to simulate Wilson Twisted mass
  Lattice QCD }                                       \\
{\em Authors: K. Jansen and C. Urbach}                                                \\
{\em Program Title: tmLQCD}                                          \\
{\em Journal Reference:}                                      \\
  %Leave blank, supplied by Elsevier.
{\em Catalogue identifier:}                                   \\
  %Leave blank, supplied by Elsevier.
{\em Licensing provisions:} GNU General Public License (GPL) \\
  %enter "none" if CPC non-profit use license is sufficient.
{\em Programming language:} C \\
{\em Computer:} any \\
  %Computer(s) for which program has been designed.
{\em Operating system:} any with a standard C compiler \\
  %Operating system(s) for which program has been designed.
{\em RAM:} no typical values available\\
%  %RAM in bytes required to execute program with typical data.
{\em Number of processors used:}                              \\
one or optionally any even number of processors. Tested with up to 32768
processors. \\
  %If more than one processor.
%{\em Supplementary material:}                                 \\
  % Fill in if necessary, otherwise leave out.
  % Please give some freely chosen keywords that we can use in a
  % cumulative keyword index.
  % see http://www.aip.org/pacs/pacs.html
{\em Classification:} 11.5 Quantum Chromodynamics, Lattice Gauge Theory \\
  %Classify using CPC Program Library Subject Index, see (
  % http://cpc.cs.qub.ac.uk/subjectIndex/SUBJECT_index.html)
  %e.g. 4.4 Feynman diagrams, 5 Computer Algebra.
{\em External routines/libraries:} LAPACK~[1] and LIME~[2] library.\\
  % Fill in if necessary, otherwise leave out.

{\em Nature of problem:}\\
Quantum Chromodynamics.
  %Describe the nature of the problem here.
   \\
{\em Solution method:}\\
Markov Chain Monte Carlo using the Hybrid Monte Carlo algorithm with
mass preconditioning and multiple time scales~[3]. Iterative
solver for large systems of linear equations.
  %Describe the method solution here.
   \\
{\em Restrictions:}\\
Restricted to an even number of (not necessarily mass degenerate)
quark flavours in the Wilson or Wilson
twisted mass formulation of lattice QCD.
  %Describe any restrictions on the complexity of the problem here.
   \\
{\em Additional comments:}\\
none.
  %Provide any additional comments here.
   \\
{\em Running time:}\\
Depending on the problem size, the architecture and the input
parameters from a few minutes to weeks.
  %Give an indication of the typical running time here.
   \\
{\em References:}
\begin{refnummer}
\item \href{http://www.netlib.org/lapack/}{{\tt
  http://www.netlib.org/lapack/}}.
% Type references in text as [1], [2], etc.
\item USQCD,
  \newblock
  \href{http://usqcd.jlab.org/usqcd-docs/c-lime/}{{\tt
  http://usqcd.jlab.org/usqcd-docs/c-lime/}}.
% This list is different from the bibliography, which
\item C.~Urbach, K.~Jansen, A.~Shindler and U.~Wenger,
  \newblock Comput. Phys. Commun. {\bf 174}, 87 (2006).
  % This is the reference list of the Program Summary
% you can use in the Long Write-Up.
\end{refnummer}
\end{small}

\newpage

% In program descriptions the main text of the paper is listed under
% the heading LONG WRITE-UP.

\hspace{1pc}
{\bf LONG WRITE-UP}

\section{Introduction}

This contribution to the anniversary issue of CPC deals with
the strong force in Particle Physics. The strong force is
presumably the least well understood fundamental interaction
between elementary particles.
It is responsible for the existence of protons and neutrons, or more
generally all nuclei, as bound states. The constituents of the nuclei 
are the quarks and gluons as the fundamental particles.
It is interesting to observe that the energy (mass) of a proton
has a size of about 1GeV while the mass of the two constituent
up and down quarks is at the order of only a few MeV. Hence, the by
far biggest contribution to the proton mass is pure binding energy.

This shows already that a description of the proton in terms of the
underlying quark and gluon degrees of freedom must be highly non-trivial.
The model that is believed to provide a theoretical framework
for the strong interaction and should give such a description
is Quantum Chromodynamics (QCD).
Although this theory can be written in a very compact mathematical
form, it is a highly non-linear theory that does not allow
for a closed analytical solution.

However, and rather fortunately, QCD can be reformulated in such
a way that computational physics methods can be applied to calculate
properties of QCD from first principles and without relying on
approximations.
In this approach, space and time are rendered discrete and a lattice
spacing $a$ is introduced. Thus, a 4-dimensional space-time lattice
is considered and the quark and gluon degrees of freedom are placed
on the lattice points or on so-called links that connect lattice points.
In this way one obtains a model of ``quark'' spins, which are coupled
to nearest neighbours only, very much reminiscent of an Ising
model in statistical physics. Indeed, the methods of statistical
physics, namely the evaluation of the partition function by means of
numerical simulations using Monte Carlo methods employing importance
sampling, are the key to address QCD on the 4-dimensional lattice, which
we will refer to as lattice QCD (LQCD).

Although the concepts to treat LQCD numerically are very clear, the problem
has an intrinsically extremely high computational demand. The crucial factor
is
that in the end the introduced discretisation has to be removed again.
If we consider a lattice with say, $L=3\ \mathrm{fm}$ linear extent and a lattice
spacing of $a=0.1\ \mathrm{fm}$ then we would have to use $N=3/0.1=30$ lattice points
in one direction. Since we are dealing with a 4-dimensional problem, we need
hence $30^4$ lattice points for such a more or less reasonable physical
situation. Simulations on such a lattice require in LQCD already several
Teraflops. Keeping $L=3\ \mathrm{fm}$ fixed and decreasing the lattice spacing to
remove the discretisation by,
say, a factor of two increases the cost of the simulation already by a factor
$2^4$. As this would not be worse enough, the used algorithms contribute
another factor of $2^{(2-3)}$. Hence, going to really fine discretisations
where the effects of the non-zero lattice spacing  
can safely be neglected or at least a controlled
extrapolation to zero values of the lattice spacing can be performed is an
extremely demanding computational challenge which will finally require
at least Petaflops computing, an area of computing power we are
realising today. 

\begin{figure}[t]
  \centering
  \subfigure[]
  {\label{fig:berlin-walla} \includegraphics[width=.443\linewidth]
    {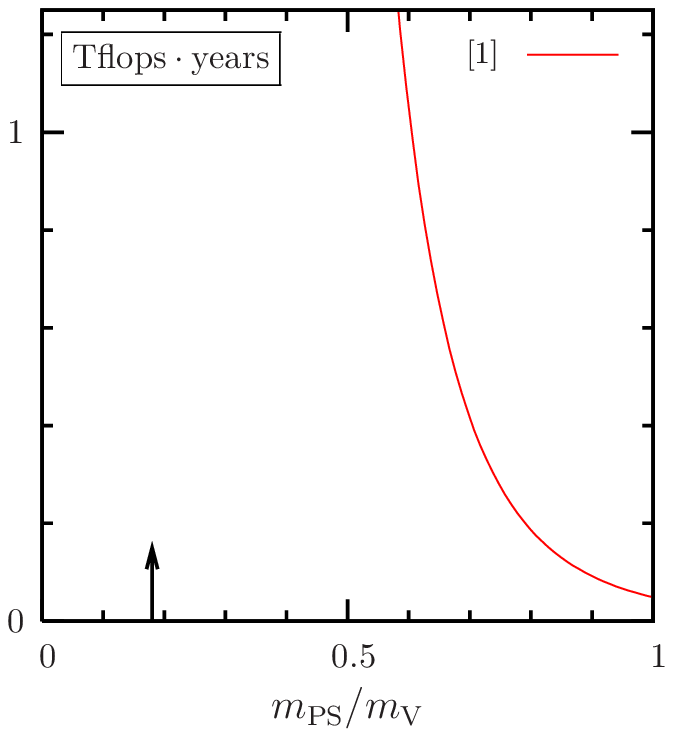}}
  \subfigure[]
  {\label{fig:berlin-wallb} \includegraphics[width=.45\linewidth]
    {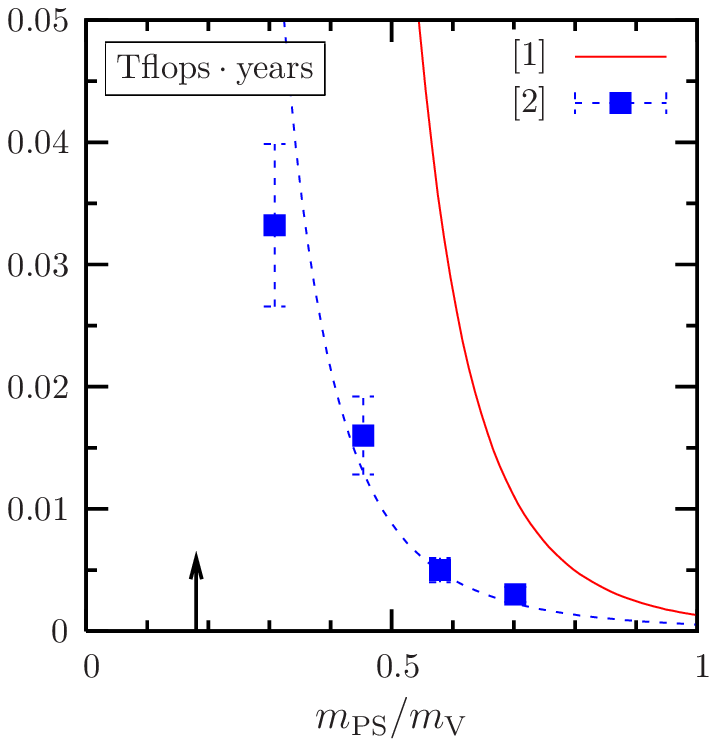}}
  \caption[Computer resources as a function of the quark mass ``Berlin
  Wall'' figure.]
  {Computer resources needed to generate $1000$ independent
    configurations of size $24^3\times 40$ at a lattice spacing of about
    $0.08\ \mathrm{fm}$ with pure Wilson fermions in units of
    $\mathrm{Tflops}\cdot\mathrm{years}$ as a function of
    $m_\mathrm{PS}/m_\mathrm{V}$. In (a) we show the prediction of
    Ref.~\cite{Ukawa:2002pc} from the  2001 lattice conference in
    Berlin (hence titled ``Berlin Wall'').
    In (b) we compare to the formula of Ref.~\cite{Ukawa:2002pc}
    (solid line) with the performance of the algorithm described in
    this paper and first published in Ref.~\cite{Urbach:2005ji}. The
    dashed line is to guide the eye. The scale of the vertical axis changes
    by about a factor of $1/50$ from (a) to (b). In (a) and (b) the
    arrow indicates the physical pion to rho meson mass ratio.
    Note that all the cost data were scaled to match a lattice
    time extend of $T=40$.}
  \label{fig:wall}
\end{figure}

However, even with the advent of Petaflops computers, the goal of
``solving'' QCD on a lattice would be completely out of reach without some
algorithmic improvements that were invented in recent years. 
This is shown in fig.~\ref{fig:wall}. In the left panel of the graph, the 
number of Teraflops years for a certain typical simulation 
is shown as a function of the ratio of two meson masses, the pseudo scalar
and the vector meson. The graph derives from the known cost 
of the used algorithm in the year 2001~\cite{Ukawa:2002pc}. 

It is important to realise that 
the meson mass ratio used in fig.~\ref{fig:wall}
assumes a value of about 0.2 in nature as 
measured in experiments, which is indicated by the arrow in both
panels of fig.~\ref{fig:wall}. The figure clearly demonstrates the strongly
growing cost of the simulations when the real physical situation is to be reached. 
In fact, simulations directly at the physical value of this mass ratio were 
impossible in 2001.
The right panel of the graph demonstrates the change of the situation 
when algorithmic improvements are used as the ones described in this 
article. In fact, the simulation costs shown in the right panel 
of fig.~\ref{fig:wall} originate from direct 
performance measurements of the code that is described here. As the 
figure demonstrates, although the simulations at the physical value 
of the meson mass ratio are still rather demanding, 
they become clearly realistic with todays Petaflops systems. There are
other approaches to improve the HMC algorithm with similar
results~\cite{Luscher:2004rx,Clark:2006fx,Clark:2006wp,AliKhan:2003mu,AliKhan:2003br,Kamleh:2005wg}.
Very promising is the recent additional improvement using inexact
deflation presented in Ref.~\cite{Luscher:2007es}.

The algorithmic improvements provided therefore a tremendous gain opening a
way for simulations in LQCD that were unthinkable a few years ago.
It is precisely the goal of this contribution to describe one programme
version of the underlying Hybrid Monte Carlo (HMC) algorithm where a number
of such improvements have been incorporated and to make the corresponding
code publicly available. The current version of the code and future
updates can be downloaded from the web~\cite{hmc:web}.

%%% Local Variables: 
%%% mode: latex
%%% TeX-master: "main"
%%% End: 

\section{Theoretical background}

\subsection{QCD on a lattice}

Quantum Chromodynamics on a hyper-cubic Euclidean space-time lattice
of size $L^3\times T$ with lattice spacing $a$ is formally described
by the action
\begin{equation}
  \label{eq:action}
  S = S_\mathrm{G}[U] + a^4 \sum_x \bar\psi\ D[U]\ \psi
\end{equation}
with $S_\mathrm{G}$ some suitable discretisation of the the Yang-Mills
action $F_{\mu\nu}^2/4$~\cite{Yang:1954ek}. The particular
implementation we are  using can be found below in section 4.2 and
consists of  plaquette and rectangular shaped Wilson loops with
particular  coefficients. $D$ is a discretisation of the Dirac
operator, for which Wilson originally proposed~\cite{Wilson:1974sk} to
use the 
so called Wilson Dirac operator
\begin{equation}
  \label{eq:DW}
  D_W[U] = \frac{1}{2}\left[\gamma_\mu\left(\nabla_\mu +
    \nabla^*_\mu\right) -a\nabla^*_\mu\nabla_\mu \right]
\end{equation}
with $\nabla_\mu$ and $\nabla_\mu^*$
the forward and backward gauge covariant difference operators,
respectively:
\begin{equation}
  \label{eq:covariant}
  \begin{split}
  \nabla_\mu\psi(x) &= \frac{1}{a}\Bigl[U(x,x+a\hat \mu)\psi(x+a \hat \mu) -
  \psi(x)\Bigr]\, , \\    
  \nabla_\mu^* \psi(x) &=
  \frac{1}{a}\Bigl[\psi(x)-U^\dagger(x,x-a\hat\mu)\psi(x-a\hat\mu)\Bigr]\, ,\\
  \end{split}
\end{equation}
where we denote the $\mathrm{SU}(3)$ link variables by $U_{x,\mu}$.
We shall set $a\equiv 1$ in the following for convenience. 
Discretising the theory is by far not a unique procedure. Instead of Wilson's
original formulation one may equally well chose the 
Wilson twisted mass formulation and the corresponding Dirac
operator~\cite{Frezzotti:2000nk}
\begin{equation}
  \label{eq:Dtm}
  D_\mathrm{tm} = (D_W[U] + m_0)\ 1_f + i \mu_q\gamma_5\tau^3
\end{equation}
for a mass degenerate doublet of quarks. We denote by $m_0$ the bare
(Wilson) quark mass, $\mu_q$ is the bare twisted
mass parameter, $\tau^i$ the $i$-th Pauli matrix and $1_f$ the
unit matrix acting in flavour space (see appendix~\ref{sec:gammas} for
our convention). In the framework of Wilson twisted mass QCD only
flavour doublets of quarks can be simulated, however, the two quarks
do not need to be degenerate in mass. The corresponding mass
non-degenerate flavour doublet reads~\cite{Frezzotti:2003xj}
\begin{equation}
  \label{eq:Dh}
  D_h(\bar\mu, \bar\epsilon)  = D_\mathrm{W}\ 1_f +
  i\bar\mu\gamma_5\tau^3 + \bar\epsilon \tau^1 \, .
\end{equation}
Note that this notation is not unique. Equivalently -- as used in
Ref.~\cite{Chiarappa:2006ae} -- one may write
\begin{equation}
  \label{eq:altDh}
  D_h'(\mu_\sigma,\mu_\delta) = D_\mathrm{W}\cdot 1_f +
  i\gamma_5\mu_\sigma\tau^1 + \mu_\delta \tau^3\, ,
\end{equation}
which is related to $D_h$ by $D_h' = (1+i\tau^2)D_h(1-i\tau^2)/2$
and $(\mu_\sigma,\mu_\delta)\to(\bar\mu, -\bar\epsilon)$. 

%%% Local Variables: 
%%% mode: latex
%%% TeX-master: "main"
%%% End: 

\subsection{The Hybrid Monte Carlo Algorithm}

For the purpose of introducing the Hybrid Monte Carlo (HMC) algorithm
we shall consider only the Wilson twisted mass formulation of 
lattice QCD with \emph{one}q doublet of mass degenerate quarks with bare quark
mass $m_0$ and bare twisted mass $\mu_q$. The extension to more than
one flavour doublet of quarks is straightforward. The corresponding
polynomial HMC algorithm used for simulating the mass non-degenerate
flavour doublet is discussed in the following sub-section.

After integrating out the Grassmann valued fermion fields, in lattice
QCD one needs to evaluate the integral
\begin{equation}
  \label{eq:partition}
  \int\ \mathcal{D}U\ \det(Q^\dagger Q)\ e^{-S_\mathrm{G}}\ ,
\end{equation}
by Markov Chain Monte Carlo methods with some discretisation of the
Yang-Mills gauge action $S_\mathrm{G}$ and  
\begin{equation}
  \label{eq:Qsq}
  Q \equiv \gamma_5 D_\mathrm{W}[U] + \gamma_5 m_0 + i\mu_q\ ,
\end{equation}
with the Wilson-Dirac operator $D_\mathrm{W}$ of eq.~(\ref{eq:DW}). 
Note that $Q$ acts now on one flavour only. The
determinant can be re-expressed using complex valued, so-called pseudo
fermion fields $\phi$ and $\phi^\dagger$
\begin{equation}
  \label{eq:det}
  \det(Q^2)\quad \propto \quad \int \mathcal{D}\phi\
  \mathcal{D}\phi^\dagger\ e^{-(Q^{-1}\phi, Q^{-1}\phi)}
\end{equation}
where $S_\mathrm{PF}\equiv |Q^{-1}\phi|^2$ is called the pseudo fermion
action. The HMC algorithm~\cite{Duane:1987de}  is then defined by 
introducing traceless hermitian momenta $P_{x,\mu}$ (conjugate to the
fundamental fields $U_{x,\mu}$) and a Hamiltonian
\begin{equation}
  \label{eq:H}
  \mathcal{H}(U,P) = \sum_{x,\mu} \frac{1}{2}\mathrm{Tr}[P_{x,\mu}^2]
  + S_\mathrm{G}[U] + S_\mathrm{PF}[U]\ .
\end{equation}
Given $\mathcal{H}$, the algorithm is composed out of a molecular
dynamics update of the fields $(U,P)\to(U',P')$ and a Metropolis
accept/reject step with respect to $\mathcal{H}$ using the acceptance 
probability 
\begin{equation}
  \label{eq:metropolis}
  P_\mathrm{acc} = \min(1, \exp\left(\mathcal{H}(U',P') -
    \mathcal{H}(U,P)\right)\, .
\end{equation}
While the momenta $P$ are generated at the beginning of a trajectory
-- in the so called heat-bath step -- randomly from a Gaussian
distribution, the pseudo fermion fields $\phi$ are generated by first
generating random fields $r$ and then
\[
\phi = Q r
\]
such that $\exp\{-(Q^{-1}\phi, Q^{-1}\phi)\} = \exp\{r^\dagger
r\}$. Note that the pseudo fermion fields are not evolved during the
molecular dynamics part of the HMC algorithm. 

\subsubsection{Molecular Dynamics Update}

In the molecular dynamics (MD) part of the HMC algorithm the momenta
and gauge fields are updated corresponding to the Hamiltonian
equations of motion
\begin{equation}
  \label{eq:hamiltoneq}
  \begin{split}
    \frac{d}{d\tau} P_{x,\mu} &= - F(x,\mu)\, ,\\
    \frac{d}{d\tau} U_{x,\mu} &= P_{x,\mu} U_{x,\mu}\, \\
  \end{split}
\end{equation}
with respect to a fictitious computer time $\tau$ and forces $F$ which
are obtained by differentiating the action with respect to the gauge
fields $U$, and takes values in the Lie algebra of
$\mathrm{SU}(3)$. The differentiation $D_U$ of some function $f(U)$ is
defined as
\[
D_U^a f(U) = \frac{\partial}{\partial\alpha} f(e^{i\alpha t^a} U)|_{\alpha=0}\, ,
\]
where $t^a$ are the generators of $\mathrm{su}(3)$.

Since these equations can in general not be integrated analytically,
one uses numerical integration methods, which must be area preserving
and reversible. Symmetrised symplectic integrators fulfil these
requirements with the simplest example being the leap-frog
algorithm. The basic discrete update steps with integration step size
$\dtau$ of the gauge field and the momenta can be defined as
\begin{equation}
  \label{eq:leapfrog}
  \begin{split}
    T_\mathrm{U}(\dtau)&:\quad U\quad\to\quad U' = \exp\left( i\dtau P\right)
      U\, ,\\
    T_\mathrm{S}(\dtau)&:\quad P\quad\to\quad P' = P - i\dtau F\, .\\
  \end{split}
\end{equation}
The leap-frog algorithm is then obtained by sequential application of 
\[
T = T_\mathrm{S}(\dtau/2)\ T_\mathrm{U}(\dtau)\
T_\mathrm{S}(\dtau/2)\, ,
\]
i.e. for a trajectory of length $\tau$ one needs to apply
$T^{N_\mathrm{MD}}$ with $N_\mathrm{MD} = \tau/\dtau$.

\subsubsection{Preconditioning and Multiple Time Scales}

Preconditioning is usually performed by factorising
\begin{equation}
  \label{eq:fact}
  \det(Q^\dagger Q) = \det(R_1^\dagger R_1)\cdot\det(R_2^\dagger
  R_2)\cdots\det(R_n^\dagger R_n)
\end{equation}
with suitably chosen $R_1, R_2, \ldots R_n$. Then For every $R_i$ a
separate pseudo fermion field $\phi_i$ is introduced, such that the
Hamiltonian reads
\begin{equation}
  \label{eq:mH}
    \mathcal{H}(U,P) = \sum_{x,\mu}
    \frac{1}{2}\mathrm{Tr}[P_{x,\mu}^2] + S_\mathrm{G}[U] + \sum_{i=1}^n
    S_{\mathrm{PF}_i}\ .
\end{equation}
and the equations of motion are changed to
\[
\begin{split}
  \frac{d}{d\tau} P_{x,\mu} &= - \sum_{i=0}^n F_i(x,\mu)\\  
  \frac{d}{d\tau} U_{x,\mu} &= P_{x,\mu} U_{x,\mu}\, \\
\end{split}
\]
where we identify $F_0$ with the force stemming from the gauge action
$S_\mathrm{G}$. 

The factorisation in eq.~(\ref{eq:fact}) can be achieved in many different
ways, see for instance
Refs.~\cite{Luscher:2004rx,Clark:2006fx,Clark:2006wp,AliKhan:2003mu,AliKhan:2003br,Kamleh:2005wg}.
Here we shall only discuss what is known as mass preconditioning or
Hasenbusch
trick~\cite{Hasenbusch:2001ne,Hasenbusch:2002ai,Hasenbusch:2003vg}. It
is obtained by writing the identity
\begin{equation}
  \label{eq:trick}
  \det(Q^\dagger Q) = \det(W^\dagger W) \cdot \frac{\det(Q^\dagger
    Q)}{\det(W^\dagger W)}\, ,
\end{equation}
where
\[
W = D_\mathrm{W} + m_0  + i \mu_2\gamma_5,\qquad \mu_2 > \mu_q\, .
\]
By adjusting the the additional mass parameter $\mu_2$, the condition
number of $W^\dagger W$ and $(Q^\dagger Q)/(W^\dagger W)$ can both be
reduced with respect to the one of $Q^\dagger Q$ alone. As argued in
Ref.~\cite{DellaMorte:2003jj}, the optimal choice leads to a condition number of
$\sqrt{k}$ for both $W^\dagger W$ and $(Q^\dagger Q)/(W^\dagger W)$,
where $k$ is the condition number of $Q^\dagger Q$. A reduced
condition number leads to reduced force contributions in the MD
evolution and allows hence for larger values of $\dtau$.

It is important to notice that evaluating the force contribution
stemming from $(Q^\dagger Q)/(W^\dagger W)$ is more expensive in terms of
computer time than the evaluation of the contribution from $W^\dagger W$,
since it involves the iterative solution of $\varphi = 
(Q^\dagger Q)^{-1}\phi$ with the large condition number $k$. Thus, the
algorithm might be further improved by not tuning the condition
numbers equal as explained beforehand, but by introducing a multiple
time scale integration scheme as follows. 

Considering a Hamiltonian like in eq.~(\ref{eq:mH}) we may introduce $n+1$
timescales $\dtau_i$ with 
\[
\dtau_i = \frac{\tau}{N_{\mathrm{MD}_i}},\qquad
N_{\mathrm{MD}_i}=N_n\cdot N_{n-1}\cdots N_i
\]
and basic discrete update steps
\begin{equation}
  \label{eq:mupdate}
  \begin{split}
    T_\mathrm{U}(\dtau)&:\quad U\quad\to\quad U' = \exp\left( i\dtau
      P\right) U\, ,\\
    T_{\mathrm{S}_i}(\dtau)&:\quad P\quad\to\quad P' = P - i\dtau 
    F_i\, \\  
  \end{split}
\end{equation}
with $0\leq i\leq n$. We have identified $S_0\equiv S_\mathrm{G}$. The
leap frog update on timescale $i$ is then recursively defined as
\begin{equation}
  \label{eq:Ti}
  T_i =
  \begin{cases}
    \Bigr.\bigl.T_{\mathrm{S}_i}(\dtau_i/2) T_\mathrm{U}(\dtau_i)\
    T_{\mathrm{S}_i}(\dtau_i/2) & i=0 \\
    \Bigr.\bigl.T_{\mathrm{S}_i}(\dtau_i/2)\ [T_{i-1}]^{N_{i-1}}\
    T_{\mathrm{S}_i}(\dtau_i/2) & 0 < i \leq n\\
  \end{cases}
\end{equation}
and the full trajectory of length $\tau$ is eventually achieved by
$[T_n]^{N_n}$. 

As was shown in Ref.~\cite{Urbach:2005ji} -- and for other
factorisations of the determinant in
Refs.~\cite{Luscher:2004rx,AliKhan:2003br,Peardon:2002wb} -- the
combination 
of multiple time scale integration and a determinant factorisation
allows to set the algorithm up such that the most expensive operator
contributes least to the MD forces. It can then be integrated on the
outermost timescale and must be less often inverted.

\subsubsection{Integration Schemes}

During the last paragraphs we have introduced the simplest reversible
and area preserving integration scheme, known as leap frog integration
scheme. There are more involved integration schemes available, partly
or completely cancelling higher order discretisation errors. 

It turns out that completely cancelling higher order effects is not
necessary and often even not efficient. Integration schemes with reduced
errors are for example the so called $n$-th order minimal norm
integration schemes, for details see Ref.~\cite{Takaishi:2005tz} and
references therein. The second order minimal norm (2MN) integration
scheme is based on the update step 
\begin{equation}
  \label{eq:2mn}
  \begin{split}
    \Bigl.\Bigr.T_{0}^\mathrm{2MN}\ =&\ T_{\mathrm{S}_0}(\lambda_0\dtau_0)\ 
    T_\mathrm{U}(\dtau_0/2)\ T_{\mathrm{S}_0}((1-2\lambda_0)\dtau_0)\ \\
    & T_\mathrm{U}(\dtau_0/2)\ T_{\mathrm{S}_0}(\lambda_0\dtau_0),  \\
    \Bigl.\Bigr.T_{i}^\mathrm{2MN}\ =&\ T_{\mathrm{S}_i}(\lambda_i\dtau_i)\ 
    [T_{i-1}^\mathrm{2MN}]^{N_{i-1}}\ T_{\mathrm{S}_i}((1-2\lambda_i)\dtau_i)\ \\
    & [T_{i-1}^\mathrm{2MN}]^{N_{i-1}}\ T_{\mathrm{S}_i}(\lambda_i\dtau_i),  \\
  \end{split}
\end{equation}
$\lambda_i$ is a dimensionless parameter and the 2MN scheme coincides with
the Sexton-Weingarten scheme~\cite{Sexton:1992nu} in case
$\lambda_i=1/6$. The optimal value for $\lambda_i$ was given in
Ref.~\cite{Takaishi:2005tz} to be around $0.19$. But its value is
likely to depend on the mass values and the time scale under
consideration. Note that there is a parameter $\lambda_i$ for each timescale
$\dtau_i$, which can be tuned separately.

While all the integration schemes introduced so far were based on the
order $T_S\ T_U\ T_S$, it is also possible to revert this order. In
this case one talks about the \emph{position} version of the
corresponding integration scheme, while the usual one is called the
\emph{velocity} version. Under certain circumstances they can be more
efficient, because one less application of $T_S$ is needed.
The corresponding update steps can be easily derived from the formulae
provided above.

\subsection{Polynomial HMC for a non-degenerate doublet}
\label{sec:ndphmc}

In the framework of Wilson twisted mass fermions it is only possible
to simulate flavour doublets of quarks. Hence, if one wants to include
the strange quark in the simulation one also needs to include the
charm. The corresponding mass non-degenerate doublet was defined in
equation (\ref{eq:Dh}). Simulating such a flavour doublet operator is
possible using the polynomial HMC (PHMC)
algorithm~\cite{Frezzotti:1997ym,Frezzotti:1998eu,Frezzotti:1998yp}.
The basic problem that occurs in the mass non-degenerate case is that 
a single flavour has to be taken into account or equivalently 
the determinant of a single operator $Q$ needs to be treated. The PHMC 
algorithm can solve this problem elegantly. 

The idea of the PHMC is based on writing
\[
\det(Q) = \det(\sqrt{Q^2}) \approx \det(P_{\epsilon,n}^{-1}(Q^2)) \propto \int\
\mathcal{D}\phi\ \mathcal{D}\phi^\dagger\ e^{-\phi^\dagger P\phi},
\]
valid as long as $Q$ is positive. $P_{\epsilon,n} (Q^2)$ is a
polynomial approximation of $1/\sqrt{Q^2}$ of degree $n$ in the
interval $[\epsilon,1]$
\begin{equation}
  \label{eq:P}
  P_{n,\epsilon}(s) = \frac{1}{\sqrt{s}}\{1 + R_{n,\epsilon}\},\qquad s =
    Q^2\, .
\end{equation}
$R_{n,\epsilon}$ is the error term. It can be shown that for the case
of Chebysheff polynomials $|R|$ vanishes exponentially fast with the
degree $n$ (for large $n$). For more details regarding this issue we
refer the reader for instance to
Refs.\cite{Luscher:1993xx,Bunk:1998rm} and references therein. 

It is worth noticing that representing in inverse operator by a polynomial has
conceptual advantages. It allows to treat certain regions of the eigenvalue
spectrum of the operator in different ways and to separate therefore 
the infrared from the bulk and ultraviolet parts of the spectrum. Although this
has been the main underlying idea of the PHMC algorithm 
\cite{Frezzotti:1997ym,Frezzotti:1998eu,Frezzotti:1998yp}
we will use it here, however, only as a technical tool to treat single flavours 
in the simulations.

For our purpose -- introducing $Q_h = \gamma_5 D_h$ -- we can rewrite
the corresponding determinant 
\[
\det(Q_h) \propto \int\ \mathcal{D}\Phi^\dagger\ \mathcal{D}\Phi\
e^{-\Phi^\dagger P_{n,\epsilon}(s)\Phi} \, ,
\]
with $s=Q_h^\dagger Q_h$ and the pseudo fermion fields $\Phi$ are now two
flavour fields. Note that $D_h^\dagger =
\tau^1 \gamma_5 D_h\gamma_5\tau^1$. The application of the polynomial
$P$ to a pseudo fermion field $\Phi$ can be performed by either using
the Clenshaw recursion relation~\cite{press:1992}, or by using the product
representation
\[
P_{n,\epsilon}(s)\Phi = \left[\prod_{i=1}^n c (s-z_i)\right]
\Phi\equiv B(s)\cdot B(s)^\dagger \Phi
\]
with $z_i$ the complex roots of $P$ and a suitably chosen
normalisation constant $c$. The product representation is conveniently
used in the MD update. For the choice of polynomials, the
determination of their roots and how to order them to avoid round-off
errors see appendix~\ref{sec:root}.

The HMC algorithm requires an area preserving and reversible MD update
procedure, however, there is no need to use in the MD update the same
operator as in the heat-bath step. As long as the acceptance rate is
sufficiently high, we are free to use any other operator in the
update. In order to exploit this possibility we introduce a second
more precise polynomial
\begin{equation}
  \label{eq:Ptilde}
  \tilde P_{m, \delta} (s)= \frac{1}{P_{n,\epsilon}\sqrt{s}}\{1+\tilde
  R_{m,\delta}\} 
\end{equation}
which is used in the heat-bath step to generate the pseudo fermion
fields from a random field $R$
\[
\Phi = \tilde P B^\dagger Q_h R
\]
and in the acceptance step. The less precise polynomial $P$ is then
used \emph{only} in the MD update. 

The polynomial degrees $n,m$ and the approximation intervals have to
be determined such as to guarantee a good approximation of
$1/\sqrt{s}$ in the range of eigenvalues of $Q_h^\dagger Q_h$. One may
also adopt a strategy to chose $\epsilon$ or $\delta$ larger than a
few lowest eigenvalues of $Q_h^\dagger Q_h$ and use re-weighting to
correct for this~\cite{Frezzotti:1997ym,Frezzotti:1998eu}.

\subsubsection*{Even/Odd preconditioning}

The (P)HMC algorithm is implemented using even/odd
preconditioning~\cite{DeGrand:1990dk,Jansen:1997yt}, which is
discussed shortly in appendix~\ref{sec:eo}.
We want to stress that although even/odd preconditioning is a rather technical 
step, it leads to a very important improvement of the algorithm 
performance and is a cornerstone of all HMC implementations in the field.

\subsection{Boundary Conditions}

The theory is discretised and put on a finite, hyper-cubic space-time
lattice with extensions $L^3\times T\equiv\ \prod_\mu L_\mu$. The boundary conditions for the
gauge fields $U_{x,\mu}$ are chosen to be periodic, i.e.
\[
U_{x+L_\nu\hat\nu,\mu} = U_{x,\mu}\ ,
\]
where $\hat\nu$ is a unit vector in direction $\nu$.
For the fermionic fields $\psi(x)$ we allow for more general boundary
conditions, namely so called twisted boundary conditions
\[
\psi(x + L_\nu\hat\nu)  = e^{i\theta_\nu \pi}\psi(x)\ .
\]
Periodic boundary conditions correspond to $\theta_\nu = 0$, while
anti-periodic boundary conditions are achieved by setting $\theta_\nu =
1$. More generally one can realise with twisted boundary conditions
arbitrary values of momentum transfer on the lattice by a convenient
re-interpretation of the phases~\cite{Sachrajda:2004mi}.

%%% Local Variables: 
%%% mode: latex
%%% TeX-master: "main"
%%% End: 

\section{Overview of the software structure}

The general strategy of the tmLQCD package is to provide programs for
the main applications used in lattice QCD with Wilson twisted mass
fermions. The code and the algorithms are designed to be general
enough such as to compile and run efficiently on any modern computer 
architecture. This is achieved code-wise by using standard C as
programming language and for parallelisation the message passing
interface (MPI) standard version 1.1.

Performance improvements are achieved by providing dedicated code for
certain widely used architectures, like PC's or the Blue Gene family. 
Dedicated code is mainly available for the kernel routine -- the
application of the Dirac operator, which will be discussed in
detail in section~\ref{sec:dirac}, and for the communication
routines.

The tmLQCD package provides three main applications. The first is an
implementation of the (P)HMC algorithm, the second and the third are
executables to invert the Wilson twisted mass Dirac operator
(\ref{eq:Dtm}) and the non-degenerate Wilson twisted mass Dirac operator
(\ref{eq:Dh}), respectively. All three do have a wide range of
run-time options, which can be influenced using an input file. The
syntax of the input file is explained in the documentation which ships
with the source code. The relevant input parameters will be mentioned
in the following where appropriate, to ease usage.

We shall firstly discuss the general layout of the three
aforementioned applications, followed by a general discussion of the
parallelisation strategy used in all three of them.

\subsection{{\ttfamily hmc\_tm}}

\begin{figure}[t]
  \centering
  \includegraphics[width=0.7\linewidth]{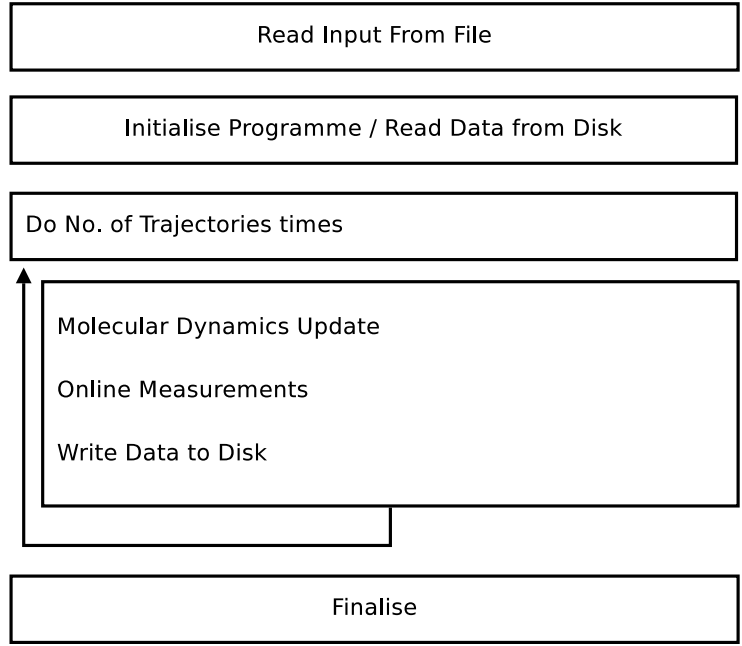}
  \caption{Flowchart for the {\ttfamily hmc\_tm} executable}
  \label{fig:hmcflow}
\end{figure}

In figure~\ref{fig:hmcflow} the programme flow of the {\ttfamily
  hmc\_tm} executable is depicted. In the first block the input file
is parsed and parameters are set accordingly. Then the required memory
is allocated and, depending on the input parameters, data is read from
disk in order to continue a previous run. 

The main part of this application is the molecular dynamics
update. For a number of trajectories, which must be specified in the
input file, first a heat-bath is performed, then the integration
according to the equations of motion using the integrator as specified
in the input file, and finally the acceptance step. 

After each trajectory certain online measurements are performed,
such as measuring the plaquette value. Other online measurements are
optional, like measuring the pseudo scalar correlation function. 

\subsubsection{command line arguments}

The programme offers command line options as follows:
\begin{itemize}
\item {\ttfamily -h|?} prints a help message and exits.
\item {\ttfamily -f} input file name. The default is {\ttfamily
    hmc.input}
\item {\ttfamily -o} the prefix of the output filenames. The default is
  {\ttfamily output}. The code will generate or append to two files,
  {\ttfamily output.data} and {\ttfamily output.para}.
\end{itemize}

\subsubsection{Input / Output}

The parameters of each run are read from an input file with default
name {\ttfamily hmc.input}. If it is missing all parameters will be
set to their default values. Any parameter not set in the input file
will also be set to its default value.

During the run the {\ttfamily hmc\_tm} program will generate two
output files, one called per default {\ttfamily output.data}, the
other one {\ttfamily output.para}. Into the latter important
parameters will be written at the beginning of the run.

The file {\ttfamily output.data} has several columns with the
following meanings
\begin{enumerate}
\item Plaquette value.
\item $\Delta H$
\item $\exp(-\Delta H)$
\item a pair of integers for each pseudo fermion monomial. The first
  integer of each pair is the sum of solver iterations needed
  in the acceptance and heatbath steps, the second is the sum of 
  iterations needed for the force computation of the whole trajectory.
\item Acceptance ($0$ or $1$).
\item Time in seconds needed for this trajectory.
\item Value of the rectangle part in the gauge action, if used.
\end{enumerate}
Every new run will append its numbers to an already existing file.

In addition, the program will create a file {\ttfamily
  history\_hmc\_tm}. This file provides a mapping between the
configuration number and its plaquette and Polyakov loop
values. Moreover the simulation parameters are stored there and in
case of a reread the time point can be found there.

After every trajectory the program will save the current configuration
in the file {\ttfamily conf.save}.

\subsection{{\ttfamily invert} and {\ttfamily invert\_doublet}}

\begin{figure}[t]
  \centering
  \includegraphics[width=0.7\linewidth]{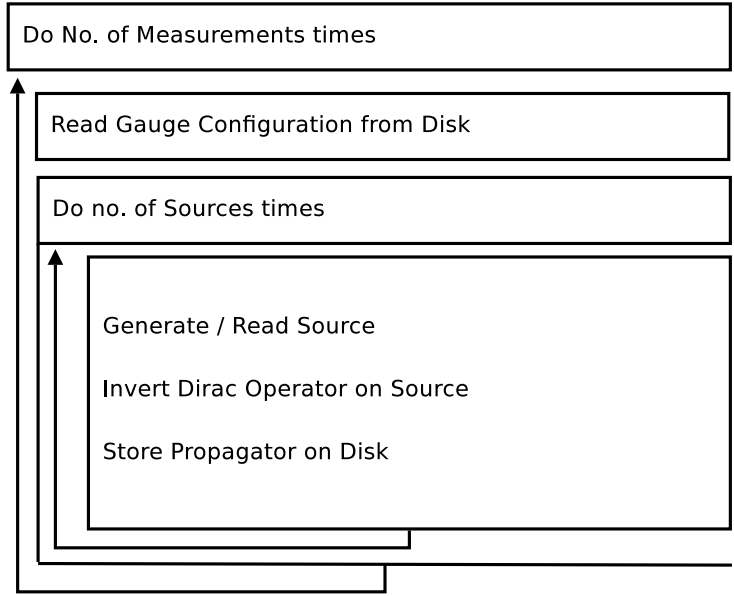}
  \caption{Flowchart for the main part of the {\ttfamily invert} and
    {\ttfamily invert\_doublet} executables.}
  \label{fig:invertflow}
\end{figure}

The two applications {\ttfamily invert} and {\ttfamily
  invert\_doublet} are very similar. The main difference is that in
{\ttfamily invert} the one flavour Wilson twisted mass Dirac operator
is inverted, whereas in {\ttfamily invert\_doublet} the non-degenerate
doublet is inverted. 

The main part of the two executables is depicted in
figure~\ref{fig:invertflow}. Each measurement corresponds to one gauge
configuration that is read from disk into memory. For each of these
gauge configurations a number of inversions will be performed. 

The sources can be either generated or read in from disk. In
the former case the programme can currently generate point sources at
random location in space time. In the latter case the name of the
source file can be specified in the input file. 

The relevant Dirac operator is then inverted on each source and the
result is stored on disk. The inversion can be performed with a number
of inversion algorithms, such as conjugate gradient (CG), BiCGstab,
and others~\cite{saad:2003a}. And optionally even/odd preconditioning
as described previously can be used. 

\subsubsection{command line arguments}

The two programmes offer command line options as follows:
\begin{itemize}
\item {\ttfamily -h|?} prints a help message and exits.
\item {\ttfamily -f} input file name. The default is {\ttfamily
    hmc.input}
\item {\ttfamily -o} the prefix of the output filenames. The default is
  {\ttfamily output}. The code will generate or append to one file
  called {\ttfamily output.para}.
\end{itemize}

\subsubsection{Output}

The program will create a file called {\ttfamily output.data} with
information about the parameters of the run. 
Of course, also the propagators are stored on disk. The corresponding
file names can be influenced via input parameters. The file format
is discussed in some detail in sub-section~\ref{sec:io}.

One particularity of the {\ttfamily invert\_doublet} program is that
the propagators written to disk correspond to the two flavour Dirac
operator of eq.~(\ref{eq:altDh}), i.e.
\[
D_h'(\mu_\sigma,\mu_\delta) = D_\mathrm{W}\cdot 1_f +
i\mu_\sigma\tau^1 + \gamma_5 \mu_\delta \tau^3\, ,
\]
essentially for compatibility reasons. For the two flavour components
written the first is the would be \emph{strange} component and the
second one the would be \emph{charm} one.                             

\subsection{Parallelisation}

The whole lattice can be parallelised in up to 4 space-time directions.
It is controlled with configure switches, see section~\ref{sec:config}.
The Message Passing Interface (MPI, standard version 1.1)  is used to
implement the parallelisation. So for compiling the parallel
executables a working MPI implementation is needed.

Depending on the number of parallelised space-time directions the
$t$-direction, the $t$- and $x$-direction, the $t$-, $x$- and
$y$-direction or the $t$-, $x$- and $y$- and $z$-direction are
parallelised. 

The number of processors per space direction must be specified at run time,
i.e. in the input file. The relevant parameters are {\ttfamily
  NrXProcs}, {\ttfamily NrYProcs} and {\ttfamily NrZProcs}. The number
of processors in time direction is determined by the program
automatically. Note that the extension in any direction must divide by
the number of processors in this direction. 

In case of even/odd preconditioning further constraints have to be
fulfilled: the local $L_z$ and the local product $L_t\times L_x \times
L_y$ must both be even.

\begin{figure}[htbp]
\centering
\includegraphics[width=0.65\linewidth]{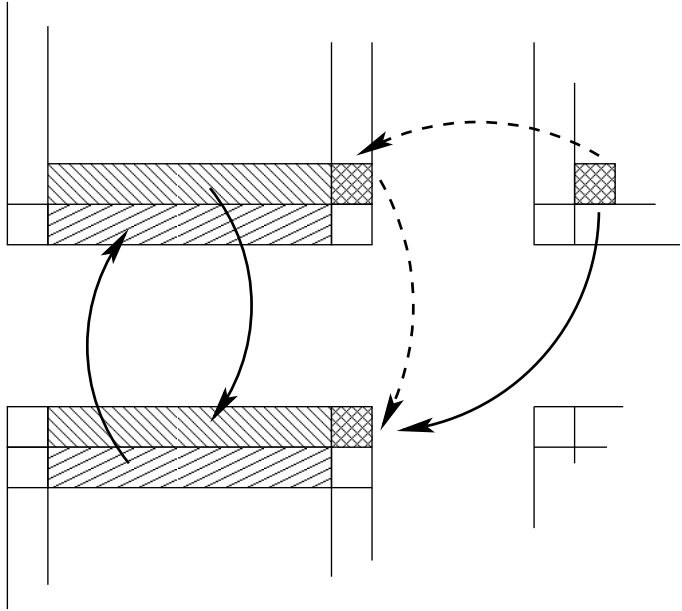}
\caption{Boundary exchange in a two dimensional parallel setup. One
  can see that the internal boundary is send while the external one
  is received. The corners -- needed for implementing improved gauge
  actions like the tree-level Symanzik improved gauge action
  \protect{\cite{Weisz:1982zw}} -- need a two step procedure.} 
\label{fig:partition}
\end{figure}

The communication is organised using boundary buffer, as sketched in
figure~\ref{fig:partition}. 
%In general the order for gauge and spinor fields in memory is as
%follows: first the local fields, then the \emph{right} t-boundary fields, then
%the \emph{left} t-boundary fields, then the \emph{right} x-boundary
%fields, the \emph{left} x-boundary fields and finally the corners (see
%figure \ref{fig:partition}). 
%
The MPI setup is contained in the file {\ttfamily mpi\_init.c}. The
corresponding function must be called at the beginning of a main
program just after the parameters are read in, also in case of a
serial run. In this function also 
the various {\ttfamily MPI\_Datatype}s are constructed needed for the
exchange of the boundary fields. The routines performing the
communication for the various data types are located in files starting
with {\ttfamily xchange\_}.

The communication is implemented using different types of MPI
functions. One implementation uses the {\ttfamily MPI\_Sendrecv}
function to communicate the data. A second one uses non-blocking MPI
functions and a third one persistent MPI calls. See the MPI standard
for details~\cite{mpi:web}. On machines with network capable of
sending in several directions in parallel the non-blocking version is
the most efficient one. The relevant configure switches are {\ttfamily
  --with-nonblockingmpi} and {\ttfamily --with-persistentmpi}, the
latter of which is only available for the Dirac operator with
halfspinor fields, see section~\ref{sec:dirac}.

%%% Local Variables: 
%%% mode: latex
%%% TeX-master: "main"
%%% End: 

\section{Description of the individual software components}

\subsection{Dirac Operator}
\label{sec:dirac}

The Dirac operator is the kernel routine of any lattice QCD
application, because its inverse is needed for the HMC update
procedure and also for computing correlation functions. The inversion
is usually performed by means of iterative solvers, like the conjugate
gradient algorithm, and hence the repeated application of the Dirac
operator to a spinor field is needed. Thus the optimisation of this
routine deserves special attention.

At some space-time point $x$ the application of a Wilson type Dirac
operator is mainly given by
\begin{equation}
  \label{eq:Dpsi}
  \begin{split}
    \phi(x) = & (m_0 + 4r +i\mu_q\gamma_5)\psi(x) \\
    &- \frac{1}{2}\sum_{\mu = 1}^4\Bigl[
    U_{x,\mu}(r+\gamma_\mu) \psi(x+a\hat\mu)  + U^\dagger_{x-a\hat\mu,\mu}
    (r-\gamma_\mu)\psi(x-a\hat\mu)\Bigr] \\
  \end{split}
\end{equation}
where $r$ is the Wilson parameter, which we set to one in the
following. The most computer time consuming part is the nearest
neighbour interaction part.

For this part it is useful to observe that 
\[
(1\pm \gamma_\mu)\psi
\]
has only two independent spinor components, the other two follow
trivially. So only two of the components need to be computed, then
to be multiplied with the corresponding gauge field $U$, and then the
other two components are to be reconstructed.

The operation in eq.~(\ref{eq:Dpsi}) must be performed for each space-time
point $x$. If the loop over $x$ is performed such that all elements
of $\phi$ are accessed sequentially (one output stream), it is clear
that the elements in $\psi$ and $U$ cannot be accessed sequentially as
well. This non-sequential access may lead to serious performance
degradations due to too many cache misses, because modern processing
units have only a very limited number of input streams available. 

While the $\psi$ field is usually different from
one to the next application of the Dirac operator, the gauge field
stays often the same for a large number of applications. This is for
instance so in iterative solvers, where the Dirac operator is applied
$\mathcal{O}(1000)$ times with fixed gauge fields. Therefore it is
useful to construct a double copy of the original gauge field sorted
such that the elements are accessed exactly in the order needed in the
Dirac operator. For the price of additional memory, with this simple
change one can obtain large performance improvements, depending on the
architecture. The double copy must be updated whenever the gauge field
change. This feature is available in the code at configure time, the
relevant switch is {\ttfamily --with-gaugecopy}.

Above we were assuming that we run sequentially through the resulting
spinor field $\phi$. Another possibility is to run sequentially
through the source spinor field $\psi$. Moreover, one could split up
the operation (\ref{eq:Dpsi}) following the standard trick of
introducing intermediate result vectors $\varphi^\pm$ with only two
spinor components per lattice site. Concentrating on the hopping part
only, we would have
\begin{equation}
  \label{eq:Dsplit}
  \begin{split}
    \varphi^+(x, \mu) &= P_{+\mu}^{4\to2}\ U_{x,\mu}(r+\gamma_\mu) \psi(x) \\
    \varphi^-(x, \mu) &= P_{-\mu}^{4\to2}\ (r-\gamma_\mu) \psi(x)\; . \\
  \end{split}
\end{equation}
From $\varphi^\pm$ we can then reconstruct the resulting spinor field 
as
\begin{equation}
  \label{eq:Dunsplit}
  \begin{split}
    \phi(x) =& \sum_\mu P_{+\mu}^{2\to4}\varphi^+(x+a\hat\mu,\mu) \\
    & + \sum_\mu P_{-\mu}^{2\to4}U^\dagger_{x-a\hat\mu,\mu}\varphi^-(x-a\hat\mu,\mu)
%    \phi(x-a\hat\mu) &= P_{+\mu}^{2\to4}\ \varphi^+(x, \mu) \\
%    \phi(x+a\hat\mu) &= P_{-\mu}^{2\to4}\
%    U^\dagger_{x-a\hat\mu,\mu}\varphi^-(x, \mu\; .)
  \end{split}
\end{equation}
Here we denote with $P_{\pm\mu}^{4\to2}$ the projection to the two
independent spinor components for $1\pm\gamma_\mu$ and with
$P_{\pm\mu}^{2\to4}$ the corresponding reconstruction.
The half spinor fields $\varphi^\pm$ can be interlaced in
memory such that $\psi(x)$ as well as $\varphi^\pm(x)$ are always
accessed sequentially in memory. The same is possible for the gauge
fields, as explained above. So only for $\phi$ we cannot avoid strided
access. So far we have only introduced extra fields $\varphi^\pm$,
which need to be loaded and stored from and to main memory, and
divided the Dirac operator into two steps (\ref{eq:Dsplit}) and
(\ref{eq:Dunsplit}) which are very balanced with regard to memory
bandwidth and floating point operations.

The advantage of this implementation of the Dirac operator comes
in the parallel case. In step (\ref{eq:Dsplit}) we need only elements
of $\psi(x)$, which are locally available on each node. So this step
can be performed without 
any communication. In between step (\ref{eq:Dsplit}) and
(\ref{eq:Dunsplit}) one then needs to communicate part of
$\varphi^\pm$, however only half the amount is needed compared to a
communication of $\psi$. After the second step there is then no
further communication needed. Hence, one can reduce the amount of data
to be sent by a factor of two. 

There is yet another performance improvement possible with this form
of the Dirac operator, this time for the price of precision. One can
store the intermediate fields $\varphi^\pm$ with reduced precision,
e.g. in single precision when the regular spinor fields are in double
precision. This will lead to a result with reduced precision, however,
in a situation where this is not important, as for instance in the MD
update procedure, it reduces the data to be communicated by another
factor of two. And the required memory bandwidth is reduced as well.
This version of the hopping matrix (currently it is only implemented
for the hopping matrix) is available at configure time with the switch
{\ttfamily --enable-halfspinor}.

The reduced precision version (sloppy precision) is available through
the input parameter {\ttfamily UseSloppyPrecision}. It will be used in
the MD update where appropriate. Moreover, it is implemented in the CG
iterative solver following the ideas outlined in
Ref.~\cite{Chiarappa:2006hz} for the overlap operator.

The various implementations of the Dirac operator can be found in the
file {\ttfamily D\_psi.c} and -- as needed for even/odd
preconditioning -- the hopping matrix in the file {\ttfamily
  Hopping\_Matrix.c}. There are many different versions of these two
routines available, each optimised for a particular architecture,
e.g. for the Blue Gene/P double hummer processor or the streaming SIMD
extensions of modern PC processors (SSE2 and SSE3), see also
Ref.~\cite{Luscher:2001tx}. Martin L{\"u}scher has made available his
standard C and SSE/SSE2 Dirac operator~\cite{Luscher:sse} under the
GNU General Public License, which are partly included into the tmLQCD
package.

\subsubsection{Blue Gene Version}

The IBM PowerPC 450d processor used on the Blue Gene architecture
provides a dual FPU, which supports a set of SIMD operations working
on 32 special registers useful for lattice QCD. These operations can
be accessed using build in functions of the IBM XLC compiler.
The file {\ttfamily bgl.h} contains all macros relevant for the Blue
Gene version of the hopping matrix and the Dirac operator. 

\begin{algorithm}[t]
  \caption{$\varphi^+ = \kappa\, U\, P_{+0}^{4\to2}(1+\gamma_0)\psi$}
  \begin{algorithmic}[1]
    \STATE // load components of $\psi$ into registers
    \STATE \_bgl\_load\_rs0((*s).s0);
    \STATE \_bgl\_load\_rs1((*s).s1);
    \STATE \_bgl\_load\_rs2((*s).s2);
    \STATE \_bgl\_load\_rs3((*s).s3);
    \STATE // prefetch gauge field for next direction $(1+\gamma_1)$
    \STATE \_prefetch\_su3(U+1);
    \STATE // do now first $P_{+0}^{4\to2}(1+\gamma_0)\psi$
    \STATE \_bgl\_vector\_add\_rs2\_to\_rs0\_reg0();
    \STATE \_bgl\_vector\_add\_rs3\_to\_rs1\_reg1();
    \STATE //now multiply both components at once with gauge field $U$ and $\kappa$
    \STATE \_bgl\_su3\_multiply\_double((*U));
    \STATE \_bgl\_vector\_cmplx\_mul\_double(ka0); 
    \STATE // store the result
    \STATE \_bgl\_store\_reg0\_up((*phi[ix]).s0);
    \STATE \_bgl\_store\_reg1\_up((*phi[ix]).s1);
  \end{algorithmic}
  \label{alg:bluegene}
\end{algorithm}

A small fraction of half spinor version (see above) is given in
algorithm \ref{alg:bluegene}, which represents the operation
$\varphi^+ = \kappa\, U\, P_{+0}^{4\to2}(1+\gamma_0)\psi$. After
loading the components of $\psi$ into the special registers and 
prefetching the gauge field for the next direction (in this case
$1+\gamma_1$), $P_{+0}^{4\to2}(1+\gamma_0)\psi$ is performed. It is
then important to load the gauge field $U$ only once from memory to
registers and multiply both spinor components in parallel. 

Finally the result is multiplied with $\kappa$ (which inherits also a
phase factor due to the way we implement the boundary conditions, see
next sub-section) and stored in memory.

\subsubsection{Boundary Conditions}

As discussed previously, we allow for arbitrary phase factors in the
boundary conditions of the fermion fields. This is conveniently
implemented in the Dirac operator as a phase factor in the hopping
term
\[
\sum_\mu \Bigl[
    e^{i\theta_\mu \pi/L_\mu}\ U_{x,\mu}(r+\gamma_\mu)
    \psi(x+a\hat\mu)  + e^{-i\theta_\mu \pi/L_\mu}\
    U^\dagger_{x-a\hat\mu,\mu} 
    (r-\gamma_\mu)\psi(x-a\hat\mu)\Bigr]\, .
\]
The relevant input parameters are {\ttfamily ThetaT}, {\ttfamily
  ThetaX}, {\ttfamily ThetaY}, {\ttfamily ThetaZ}.

\subsection{The HMC Update}

We assume in the following that the action to be simulated can be
written as 
\[
S = S_\mathrm{G} + \sum_{i=1}^{N_\mathrm{monomials}} S_{\mathrm{PF}_i}\, ,
\]
and we call -- following the CHROMA notation~\cite{Edwards:2004sx} -- each
term in this sum a \emph{monomial}. We require that there is exactly one
gauge monomial $S_\mathrm{G}$ (which we identify with $S_0$ in the
following) and an arbitrary number of pseudo
fermion monomials $S_{\mathrm{PF}_i}$.

As a data type every monomial must known how to compute its
contribution to the initial Hamiltonian $\mathcal{H}$ at the beginning
of each trajectory in the heat-bath step. Then it must know how to
compute the derivative with respect to the gauge fields for given
gauge field and pseudo fermion field needed for the MD update. And finally
there must be a function to compute its contribution to the final
Hamiltonian $\mathcal{H}'$ as used in the acceptance step.

\begin{figure}[t]
  \centering
  \includegraphics[width=0.7\linewidth]{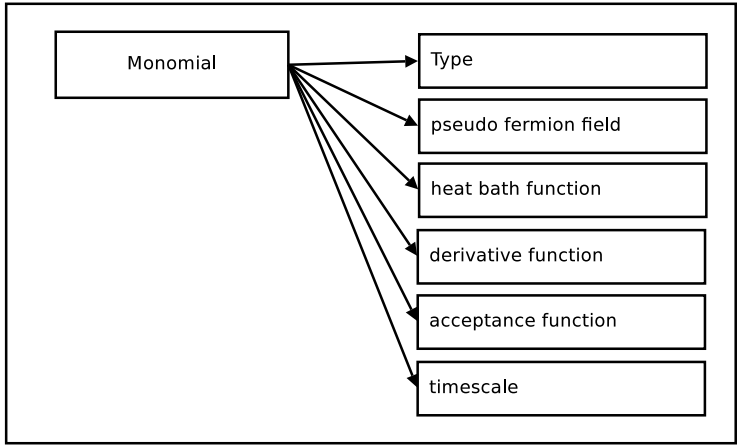}
  \caption{Data type monomial and its components}
  \label{fig:monomial}
\end{figure}

In addition for each monomial it needs to be known on which timescale
it should be integrated. The corresponding data type is sketched in
figure~\ref{fig:monomial}. The general definitions for this data type
can be found in the file {\ttfamily monomial.c}. 

There are several sorts of monomials implemented:
\begin{itemize}
\item {\ttfamily DET}: pseudo fermion representation of the (mass
  degenerate) simple determinant\\
  \[
  \det(Q^2(\kappa) + \mu^2)
  \]
\item {\ttfamily DETRATIO}: pseudo fermion representation of the
  determinant ratio\\
  \[
  \det(Q^2(\kappa) + \mu^2)/\det(Q^2(\kappa_2) + \mu_2^2)
  \]
\item {\ttfamily NDPOLY}: polynomial representation of the (possibly
  non-degenerate) doublet\\
  \[
  [\det(Q_{nd}(\bar\epsilon, \bar\mu)^2)]^{1/2}\, .
  \]
\item {\ttfamily GAUGE}:\\
  \[
  \frac{\beta}{3}\sum_x\left(  c_0\sum_{\substack{
        \mu,\nu=1\\1\leq\mu<\nu}}^4\{1-\re\tr(U^{1\times1}_{x,\mu,\nu})\}\Bigr. 
    \Bigl.\ +\ 
    c_1\sum_{\substack{\mu,\nu=1\\\mu\neq\nu}}^4\{1
    -\re\tr(U^{1\times2}_{x,\mu,\nu})\}\right)\,  ,
  \]
  The parameter $c_1$ can be set in the input file and
  $c_0=1-8c_1$. Note that $c_1=0$ corresponds to the Wilson plaquette
  gauge action. 
\end{itemize}
The corresponding specific functions are defined in the files
{\ttfamily det\_monomial.c}, {\ttfamily detratio\_monomial.c},
{\ttfamily ndpoly\_monomial.c} and {\ttfamily
  gauge\_monomial.c}. Additional monomials can easily be implemented
by providing the corresponding functions as discussed above.

\begin{algorithm}[t]
  \caption{integrate}
  \begin{algorithmic}[1]
    \REQUIRE $0 < n_\mathrm{ts}\leq N_\mathrm{ts}$, $\tau > 0$
    \STATE $\dtau = \tau/$noSteps[$n_\mathrm{ts}$]
    \FOR{$i$ = 0 to noSteps[$n_\mathrm{ts}$]}
    \IF{$n_\mathrm{ts}$ == $1$}
    \STATE updateGauge($\dtau$)
    \ELSE
    \STATE integrate($n_\mathrm{ts}-1$, $\dtau$)
    \ENDIF
    \STATE updateMomenta($\dtau$, monomialList[$n_\mathrm{ts}$])
    \ENDFOR
  \end{algorithmic}
  \label{alg:integrator}
\end{algorithm}

The integration scheme is implemented recursively, as exemplified in
algorithm~\ref{alg:integrator} for the leap-frog integration scheme
(where we skipped half steps for simplicity). The updateMomenta
function simply calls the derivative functions of all monomials
that are integrated on timescale $n_\mathrm{ts}$ and updates the
momenta $P$ according to the time step $\dtau$.

The recursive scheme for the integration can easily be extended to
more involved integration schemes. The details can be found in the
file {\ttfamily integrator.c}. We have implemented the leap-frog and
the second order minimal norm~\cite{Takaishi:2005tz} integrations
schemes. They are named in the input file as {\ttfamily LEAPFROG} and
{\ttfamily 2MN}, respectively. These two can be mixed on
different timescales. In addition we have implemented a position
version of the second order minimal norm integration scheme, denoted by
{\ttfamily 2MNPOSITION} in the input file. The latter must not be mixed with
the former two.

The MD update is summarised in
algorithm~\ref{alg:mdupdate}. It computes the initial and final
Hamiltonians and calls in between the integration function with the
total number of timescales $N_\mathrm{ts}$ and the total trajectory
length $\tau$.

\subsubsection{Reduced Precision in the MD Update}

As shortly discussed previously, as long as the integration in the MD
update is reversible and area preserving there is large freedom in
choosing the integration scheme, but also the operator: it is not
necessary to use the Dirac operator here, it can be any approximation
to it. This is only useful if the acceptance rate is not strongly
affected by such an approximation.

The code provides two possibilities to adapt the precision of the
Dirac operator used in the MD update: the first is to reduce the
precision in the inversions needed for the force computation. This
causes reduced iteration numbers needed for the integration of one
trajectory. The relevant input parameter is {\ttfamily
  ForcePrecision} available for each monomial. The precision needed in
the acceptance and/or heatbath step can be adjusted separately using
{\ttfamily AcceptancePrecision}. It is advisable to have the
acceptance precision always close to machine precision.

\begin{algorithm}[t]
  \caption{MD update}
  \begin{algorithmic}[1]
    \STATE $\mathcal{H}=\mathcal{H}'=0$
    \FOR{$i$ = 0 to $N_\mathrm{monomials}$} 
    \STATE $\mathcal{H}$ += monomial[$i$]$\rightarrow$heat-bath-function
    \ENDFOR

    \STATE integrate($N_\mathrm{ts}$, $\tau$)

    \FOR{$i$ = 0 to $N_\mathrm{monomials}$} 
    \STATE $\mathcal{H}'$ += monomial[$i$]$\rightarrow$acceptance-function
    \ENDFOR
    \STATE accept with probability $\min\{1, \exp(-\Delta\mathcal{H})\}$
  \end{algorithmic}
  \label{alg:mdupdate}
\end{algorithm}

The second possibility for influencing the Dirac operator is given by
the reduced precision Dirac operator described in
sub-section~\ref{sec:dirac}, which is switched on with the {\ttfamily
  UseSloppyPrecision} input parameter. The two possibilities can also
be used in parallel.

Note that one should always test for reversibility violations as
explained in sub-section \ref{sec:online}.

\subsubsection{Chronological Solver}

The idea of the chronological solver method, or chronological solver
guess (CSG)  (or similar methods \cite{Brower:1994er}) is to optimise
the initial guess for the solution used in the solver. To this end the
history of $N_\mathrm{CSG}$ last solutions of the equation $M^2 \chi =
\phi$ is saved and then a linear combination of the fields $\chi_i$
with coefficients $c_i$ is used as an initial guess for the next
inversion. $M$ stands for the operator to be inverted and has to be
replaced by the different ratios of operators used in this paper.

The coefficients $c_i$ are determined by solving
\begin{equation}
  \label{eq:chrono}
  \sum_i \chi_j^\dagger M^2 \chi_i c_i = \chi_j^\dagger \phi
\end{equation}
with respect to the coefficients $c_i$. This is equivalent to
minimising the functional that is minimised by the CG inverter
itself.

The downside of this method is that the reversibility violations
increase significantly by one or two orders of magnitude in the
Hamiltonian when the CSG is switched on and all other parameters are
kept fixed. Therefore one has to adjust the residues in the solvers,
which increases the number of matrix vector multiplications again.
Our experience is that the methods described in the previous
sub-section are more effective in particular in the context of
multiple time scale integration, because the CSG is most effective for
small values of $\dtau$.

The input parameters is the {\ttfamily CSGHistory} parameter
available for the relevant monomials. Setting it to zero means no
chronological solver, otherwise this parameter specifies the number of
last solutions $N_\mathrm{CSG}$ to be saved.

\subsection{Online Measurements}
\label{sec:online}

The HMC program includes the possibility to perform a certain number
of measurements after every trajectory \emph{online}, whether or not
the configuration is stored on disk. Some of those are performed per
default, namely all that are written to the output file {\ttfamily
  output.data}: 
\begin{enumerate}
\item the plaquette expectation value, defined as:
  \[
  \langle P\rangle = \frac{1}{6 V}\ \sum_{\substack{
        \mu,\nu=1\ 1\leq\mu<\nu}}^4\ \re\tr(U^{1\times1}_{x,\mu,\nu})\, ,
  \]
  where $V$ is the global lattice volume.
\item the rectangle expectation value, defined as:
  \[
  \langle R\rangle = \frac{1}{12V}\ \sum_{\substack{\mu,\nu=1\
      \mu\neq\nu}}^4\ 
    \re\tr(U^{1\times2}_{x,\mu,\nu})
  \]
\item $\Delta\mathcal{H} = \mathcal{H}'-\mathcal{H}$ and $\exp(-\Delta\mathcal{H})$.
\end{enumerate}
See the overview section for details about the {\ttfamily output.data}
file. These observables all come with no extra computational cost.

Optionally, other online measurements can be performed, which --
however -- need in general extra inversions of the Dirac
operator. First of all the computation of certain correlation
functions is implemented. They need \emph{one} extra inversion of the
Dirac operator, as discussed in Ref.~\cite{Boucaud:2008xu}, using the
one-end-trick. Define a stochastic source $\xi$ as follows
\begin{equation}
  \label{eq:source}
  \lim_{R\to\infty}[\xi_i^*\xi_j] = \delta_{ij},\quad
  \lim_{R\to\infty}[\xi_i\xi_j] = 0\, .
\end{equation}
Here $R$ labels the number of samples and $i$ all other degrees of
freedom. Then 
\begin{equation}
  \label{oneend}
  [\phi_i^{r*}\phi_j^r]_R = M_{ik}^{-1*}\cdot M_{jk}^{-1} +
  \textrm{noise}\, ,
\end{equation}
if $\phi$ was computed from
\[
\phi_j^r  = M^{-1}_{jk}\xi_k^r\, .
\]
Having in mind the $\gamma_5$-hermiticity property of the Wilson and
Wilson twisted mass Dirac propagator $G_{u,d}$, i.e.
\[
G_u(x,y) = \gamma_5 G_d(y,x)^\dagger \gamma_5
\]
it is clear that eq.~(\ref{oneend}) can be used to evaluate
\[
C_\pi(t) = \langle \tr[G_u(0,t)\gamma_5 G_d(t,0)\gamma_5]\rangle =
\langle \tr[G_u(0,t) G_u(0,t)^\dagger]\rangle
\]
with only one inversion. But, even if the one gamma structure at the
source is fixed to be $\gamma_5$ due to the $\gamma_5$-hermiticity
trick, we are still free to insert any $\gamma$-structure $\Gamma$ at the source,
i.e. we can evaluate any correlation function of the form
\[
C_{P\Gamma}(t) = \langle\tr[G_u(0,t) \gamma_5 G_d(t,0) \Gamma]\rangle
= \langle \tr[G_u(0,t) G_u(0,t)^\dagger\gamma_5\Gamma]\rangle\, .
\]
Useful combinations of correlation functions are $\langle P P\rangle$,
$\langle PA\rangle$ and $\langle PV\rangle$, with
\[
  P^\alpha = \bar\chi \gamma_5 \frac{\tau^\alpha}{2}\chi\, ,\quad
  V^\alpha_\mu = \bar\chi \gamma_\mu\frac{\tau^\alpha}{2}\chi\, ,\quad
  A^\alpha_\mu = \bar\chi \gamma_5\gamma_\mu\frac{\tau^\alpha}{2}\chi
\]
From $\langle P P\rangle$ one can extract the pseudo scalar mass, and
-- in the twisted mass case -- the pseudo scalar decay
constant. $\langle PA\rangle$ can be used together with $\langle P
P\rangle$ to extract the so called PCAC quark mass and $\langle
PV\rangle$ to measure the renormalisation constant $Z_\mathrm{V}$. For
details we refer the reader to Ref.~\cite{Boucaud:2008xu}.

These online measurements are controlled with the two following input
parameters: {\ttfamily PerformOnlineMeasurements} to switch them on or
off and to specify the frequency {\ttfamily OnlineMeasurementsFreq}. The three
correlation functions are saved in files named {\ttfamily
  onlinemeas.n}, where {\ttfamily n} is the trajectory number. Every
file contains five columns, specifying the type, the operator type and the
Euclidean time $t$. The last two columns are the values of the
correlation function itself, $C(t)$ and $C(-t)$, respectively. The
type is equal to $1$, $2$ or $6$ for the $\langle P P\rangle$, the
$\langle PA\rangle$ and the $\langle PV\rangle$ correlation
functions. The operator type is for online measurements always equal
to $1$ for local source and sink (no smearing of any kind), and the
time runs from $0$ to $T/2$. Hence, $C(-t)= C(T-t)$. $C(-0)$ and
$C(-T/2)$ are set to zero for convenience.

In addition to correlation functions also the minimal and the maximal
eigenvalues of the $(\gamma_5 D)^2$ can be measured.

An online measurement not related to physics, but related to the
algorithm are checks of reversibility violations. The HMC algorithm is
exact if and only if the integration scheme is reversible. On a
computer with finite precision this is only guaranteed up to machine
precision. These violations can be estimated by integrating one
trajectory forward and then backward in Monte Carlo time. The
difference $\delta\Delta\mathcal{H}$ among
the original Hamiltonian $\mathcal{H}$ and the final one
$\mathcal{H}''$ after integrating back can serve as one measure for
those violations, another one is provided by the difference among the
original gauge field $U$ and the final one $U''$
\[
\delta\Delta U = \frac{1}{12V}
\sum_{x,\mu}\sum_{i,j} (U_{x,\mu}-U_{x,\mu}'')_{i,j}^2
\]
where we indicate with the $\delta\Delta$ that this is obtained after
integrating a trajectory forward and backward in time. The results for
$\delta\Delta \mathcal{H}$ and $\delta\Delta U$ are
stored in the file {\ttfamily return\_check.data}. The relevant input
parameters are {\ttfamily ReversibilityCheck} and {\ttfamily
  ReversibilityCheckInterval}.

\subsection{Iterative Solver and Eigensolver}

There are several iterative solvers implemented in the tmLQCD
package for solving 
\[
D\ \chi = \phi
\]
for $\chi$. The minimal residual (MR), the conjugate gradient (CG), the
conjugate gradient squared (CGS), the generalised minimal residual
(GMRES), the generalised conjugate residual and the stabilised
bi-conjugate gradient (BiCGstab). For details regarding these
algorithms we refer to Refs.~\cite{saad:2003a,meister:1999}.

For the {\ttfamily hmc\_tm} executable only the CG and the BiCGstab
solvers are available, while all the others can be used in the
{\ttfamily invert} executables. Most of them are both available with
and without even/odd preconditioning. For a performance comparison we
refer to Ref.~\cite{Chiarappa:2004ry,Chiarappa:2006hz}.

The stopping criterion is implemented in two ways: the first is an
absolute stopping criterion, i.e. the solver is stopped when the
squared norm of the residual vector (depending on the solver this
might be the iterated residual or the real residual) fulfils
\[
\|r\|^2 < \epsilon^2\, .
\]
The second is relative to the source vector, i.e.
\[
\frac{\|r\|^2}{\|\phi\|^2} < \epsilon^2\, .
\]
The value of $\epsilon^2$ and the choice of relative or absolute precision can be
influenced via input parameters.

The reduced precision Dirac operator, as discussed in sub-section
\ref{sec:dirac}, is available for the CG solver. In the CG solver the 
full precision Dirac operator is only required at the beginning of the
CG search, because the relative size of the contribution to the
resulting vector decreases with the number of iterations. Thus, as soon
as a certain precision is achieved in the CG algorithm we can switch to
the reduced precision Dirac operator without spoiling the precision of
the final result. We switch to the lower precision operator 
at a precision of $\sqrt{\epsilon}$ in the CG search, when aiming for a
final precision of $\epsilon < 1$.

We note that in principle any combination of using reduced precision
in one of the ways described in this paper is possible. However, one
should always check that the true 
residual is as small as expected in case of an inversion and that the
reversibility violations are small in case of a HMC simulation.

The eigensolver used to compute the eigenvalues (and vectors) of
$(\gamma_5 D)^2$ is the so called Jacobi-Davidson 
method~\cite{Sleijpen:1996aa,Geus:2002}. For a discussion for the
application of this algorithm to lattice QCD we refer again to
Ref.~\cite{Chiarappa:2004ry,Chiarappa:2006hz}. 

All solver related files can be found in the sub-directory {\ttfamily
  solver}. Note that there are a few more solvers implemented which
are, however, in an experimental status.

\subsection{Stout Smearing}

Smearing techniques have become an important tool to reduce
ultraviolet fluctuations in the gauge fields. One of those techniques,
coming with the advantage of being usable in the MD update, is usually
called stout smearing~\cite{Morningstar:2003gk}. 

The $(n+1)^{\rm th}$ level of stout smeared gauge links is obtained iteratively
from the $n^{\rm th}$ level by
\begin{equation*}
  U_\mu^{(n+1)}(x)\;=\;e^{i\,Q_\mu^{(n)}(x)}\,U_\mu^{(n)}(x).
\end{equation*}
We refer to the unsmeared (``thin'') gauge field as $U_\mu\equiv
U_\mu^{(0)}$.
The ${\rm SU}(3)$ matrices $Q_\mu$ are defined via the staples $C_\mu$:
\begin{eqnarray}
  Q_\mu^{(n)}(x) &=& \frac{i}2\Big[U^{(n)}_\mu(x){C_\mu^{(n)}}^\dagger(x)
  - {\mathrm{h.c.}}\Big]\,-\,\frac{i}{6}\tr\Big[U^{(n)}_\mu(x){C_\mu^{(n)}}^\dagger(x)
  - {\mathrm{h.c.}}\Big]\,,\nonumber\\
  C_\mu^{(n)} &=& \sum_{\nu\neq\mu}\,\rho_{\mu\nu}\,
  \Big(U_\nu^{(n)}(x)U_\mu^{(n)}(x+\hat\nu){U_\nu^{(n)}}^\dagger(x+\hat\mu)
  \nonumber\\
  && \;\;\;
  +{U_\nu^{(n)}}^\dagger(x-\hat\nu)U_\mu^{(n)}(x-\hat\nu)U_\nu^{(n)}(x-\hat\nu+\hat\mu)
  \Big)\,,\nonumber
\end{eqnarray}
where in general $\rho_{\mu\nu}$ is the smearing matrix.
In the tmLQCD package we have only implemented isotropic $4$-dimensional
smearing, i.e., $\rho_{\mu\nu}=\rho$.

Currently stout smearing is only implemented for the {\ttfamily
  invert} executables. I.e. the gauge field can be stout smeared at
the beginning of an inversion. The input parameters are {\ttfamily
  UseStoutSmearing}, {\ttfamily StoutRho} and {\ttfamily
  StoutNoIterations}. 

\subsection{Random Number Generator}

The random number generator used in the code is the one proposed by
Martin L{\"u}scher and usually known under the name
RANLUX~\cite{Luscher:1993dy}. A single and double precision
implementation was made available by the author under the GNU General
Public License and can be downloaded~\cite{Luscher:ranluxweb}. For
convenience it is also included in the tmLQCD package.

\subsection{IO Formats}
\label{sec:io}

In this final subsection we specify the IO formats used to store gauge
configurations, propagators and sources to disk.

\subsubsection{Gauge Configurations}

For gauge configurations we use the International Lattice Data Grid
(ILDG) standard as specified in Ref.~\cite{ildg:web,Yoshie:2008aw}. As
the lime packaging library~\cite{lime:web} and ILDG standard allow
additional -- not required -- records 
to be stored within the file, we currently add the following two
records for convenience:
\begin{enumerate}
\item {\ttfamily xlf-info}: useful information about the gauge
  configuration, such as the plaquette value, and about the run and
  the algorithm and the code version used to generate it.

\item {\ttfamily scidac-checksum}: SCIDAC checksum of the gauge
  configuration. For the specification see~\cite{scidac}.

\end{enumerate}
The gauge configurations can be written to disk either in single or
double precision. The relevant input parameter is {\ttfamily
  GaugeConfigWritePrecision}. On readin the precision is determined
automatically. 

Note that the gauge configuration does not depend on the particular
choice of the $\gamma$-matrices.

\subsubsection{Propagators}

We note at the beginning, that we do not use different IO formats for
source or sink fermion fields. They are both stored using the same
lime records. The meta-data stored in the same lime-packed file is
supposed to clarify all other things. It is also important to realise
that the propagator depends on the $\gamma$-matrix convention used in
the Dirac operator. For our convention see appendix~\ref{sec:gammas}.

Here we mainly concentrate on storing propagators (sink). The file can
contain only sources, or both, source and sink. We (plan to) support
four different formats
\begin{enumerate}
\item (arbitrary number of) sink, no sources
\item (arbitrary number of) source/sink pairs
\item one source, 12 sink
\item one source, 4 sink
\end{enumerate}
This is very similar to the formats in use in parts of the US
lattice community. We adopt the SCIDAC checksum~\cite{scidac} for the
binary data.

Source and sink binary data has to be in a separate lime record. The
order in one file for the four formats mentioned above is supposed to
be 
\begin{enumerate}
\item sink, no sources: -
\item source/sink pairs: first source, then sink
\item one source, 12 sink: first source, then 12 sinks
\item one source, 4 sink: first source, then 4 sinks
\end{enumerate}
All fermion field files must have a record indicating its type. The
record itself is of type {\ttfamily propagator-type} and the
record has a single entry (ASCII string) which contains one of 
\begin{itemize}
\item {\ttfamily DiracFermion\_Sink}
\item {\ttfamily DiracFermion\_Source\_Sink\_Pairs}
\item {\ttfamily DiracFermion\_ScalarSource\_TwelveSink}
\item {\ttfamily DiracFermion\_ScalarSource\_FourSink}
\end{itemize}
Those strings are also used in the input files for the
input parameter {\ttfamily PropagatorType}.
The binary data corresponding to one Dirac fermion field (source or
sink) is then stored with at least two (three) records. The first is
of type \\
{\ttfamily etmc-propagator-format} \\
and contains the following information:
\begin{verbatim}
<?xml version="1.0" encoding="UTF-8"?>
<etmcFormat>
  <field>diracFermion</field>
  <precision>32</precision>
  <flavours>1</flavours>
  <lx>4</lx>
  <ly>4</ly>
  <lz>4</lz>
  <lt>4</lt>
</etmcFormat>
\end{verbatim}
The {\ttfamily flavours} entry must be set to {\ttfamily 1} for a one
flavour propagator (flavour diagonal case) and to {\ttfamily 2} for a two
flavour propagator (flavour non-diagonal 2-flavour operator). In the
former case there follows one record of type
{\ttfamily scidac-binary-data}, which is identical to the SCIDAC
format, containing the fermion field. In the latter case there follow
two of such records, the first of which is the upper flavour. To be
precise, lets call the two flavours $s$ (strange) and $c$
(charm). Then we always store the $s$ component first and then the $c$
component. 
%Any number of other records can be added for convenience.

The first two types are by now supported in the tmLQCD package. In the
future the other two might follow.

The indices (time, space, spin, colour) in the binary data {\ttfamily
  scidac-binary-data} are in the following order:
\[
t, z, y, x, s, c\, ,
\]
where $t$ is the slowest and colour the fastest running index.
The binary data is stored big endian and either in single or in double
precision, depending on the {\ttfamily precision} entry in the
{\ttfamily etmc-propagator-format} record. 

In addition we store an additional record called {\ttfamily
  inverter-info} with useful information about the inversion
precision, the physical parameters and the code version.

\subsubsection{Source Fields}

Source fields are, as mentioned before, stored with the same binary
data format. There are again several types of source files possible:
\begin{itemize}
\item {\ttfamily DiracFermion\_Source}
\item {\ttfamily DiracFermion\_ScalarSource}
\item {\ttfamily DiracFermion\_FourScalarSource}
\item {\ttfamily DiracFermion\_TwelveScalarSource}
\end{itemize}
This type is stored in a record called {\ttfamily source-type} in the
lime file. There might be several sources stored within the same
file. We add a format record {\ttfamily etmc-source-format} looking like
\begin{verbatim}
<?xml version="1.0" encoding="UTF-8"?>
<etmcFormat>
  <field>diracFermion</field>
  <precision>32</precision>
  <flavours>1</flavours>
  <lx>4</lx>
  <ly>4</ly>
  <lz>4</lz>
  <lt>4</lt>
  <spin>4</spin>
  <colour>3</colour>
</etmcFormat>
\end{verbatim}
with obvious meaning for every {\ttfamily scidac-binary-data} record
within the lime packed file. This format record also allows to store a
subset of the whole field, e.g. a time-slice.

\section{Installation instructions}

The software ships with a GNU autoconf environment and a configure
script, which will generate GNU Makefiles to build the programmes. It
is supported and recommended to configure and build the executables in
a separate build directory. This also allows to have several builds with
different options from the same source code directory. 

\subsection{Prerequisites}

In order to compile the programmes the {\ttfamily
  LAPACK}~\cite{lapack:web} library (Fortran version) needs to be
installed. In addition it must be known which linker options are
needed to link against {\ttfamily LAPACK}, e.g. {\ttfamily
  -Lpath-to-lapack -llapack  -lblas}. Also a the latest
version (tested is version 1.2.3) of {\ttfamily
  C-LIME}~\cite{lime:web} must be available, which is used as a
packaging scheme to read and write gauge configurations and
propagators to files.

\subsection{Configuring the tmLQCD package}
\label{sec:config}

In order to get a simple configuration of the hmc package it is enough
to just type 
\begin{verbatim}
patch-to-src-code/configure   --with-lime=<path-to-lime> \
     --with-lapack=<linker-flags> CC=<mycc> \
     F77=<myf77> CFLAGS=<c-compiler flags>
\end{verbatim}
in the build directory. If 
{\ttfamily CC, F77} and {\ttfamily CFLAGS} are not specified,
{\ttfamily configure} will guess them.

The code was successfully compiled and run at least on the following
platforms: i686 and compatible, x64 and compatible, IBM Regatta
systems, IBM Blue Gene/L, IBM Blue Gene/P, SGI Altix and SGI PC
clusters and powerpc clusters.

The configure script accepts certain options to influence the building
procedure. One can get an overview over all supported options with
{\ttfamily configure --help}. There are {\ttfamily enable|disable}
options switching on and off optional features and {\ttfamily
  with|without} switches usually related to optional packages. In the
following we describe the most important of them (check {\ttfamily
  configure --help} for the defaults and more options):

\begin{itemize}
\item {\ttfamily --enable-mpi}:\\
  This option switches on the support for MPI. On certain platforms it
  automatically chooses the correct parallel compiler or searches for
  a command {\ttfamily mpicc} in the search path.

\item {\ttfamily --enable-p4}:\\
  Enable the use of special Pentium4 instruction set and cache
  management.

\item {\ttfamily --enable-opteron}:\\
  Enable the use of special opteron instruction set and cache
  management.

%\item {\ttfamily --enable-sse}:\\
%  Enable the use of SSE instruction set. This means not much when 64
%  Bit precision is used.

\item {\ttfamily --enable-sse2}:\\
  Enable the use of SSE2 instruction set. This is a huge improvement
  on Pentium4 and equivalent systems.

\item {\ttfamily --enable-sse3}:\\
  Enable the use of SSE3 instruction set. This will give another 20\%
  of speedup when compared to only SSE2. However, only a few
  processors are capable of SSE3.

\item {\ttfamily --enable-gaugecopy}:\\
  See section \ref{sec:dirac} for details on this option. It will
  increase the memory requirement of the code.

\item {\ttfamily --enable-halfspinor}:\\
  If this option is enabled the Dirac operator using half spinor
  fields is used. See sub-section \ref{sec:dirac} for details. If this
  feature is switched on, also the gauge copy feature is switched
  on automatically. 

%\item {\ttfamily --enable-shmem}:\\
%  Use shared memory API instead of MPI for the communication of spinor
%  fields. This is currently only usable on the Munich Altix machine.

\item {\ttfamily --with-mpidimension=n}:\\
  This option has only effect if the code is configured for MPI usage.
  The number of parallel directions can be specified. 1,2,3 and 4
  dimensional parallelisation is supported.

\item {\ttfamily --with-lapack="<linker flags>"}:\\
  the code requires lapack to be linked. All linker flags necessary
  to do so must be specified here. Note that {\ttfamily LIBS="..."}
  works similar.

\item {\ttfamily --with-limedir=<dir>}:\\
  Tells configure where to find the lime package, which is required for
  the build of the HMC. It is used for the ILDG file format.
 
\end{itemize}

The configure script will guess at the very beginning on which
platform the build is done. In case this fails or a cross compilation
must be performed please use the option {\ttfamily --host=HOST}. For
instance in order to compile for the BG/P one needs to specify
{\ttfamily --host=ppc-ibm-bprts --build=ppc64-ibm-linux}. 

For certain architectures like the Blue Gene systems there are
{\ttfamily README.arch} files in the top source directory with
example configure calls.

\subsection{Building and Installing}

After successfully configuring the package the code can be build by
simply typing {\ttfamily make} in the build directory. This will
compile the standard executables. Typing {\ttfamily make install} will
copy these executables into the install directory. The default install
directory is {\ttfamily \$HOME/bin}, which can be influenced e.g. with
the {\ttfamily --prefix} option to {\ttfamily configure}.

%%% Local Variables: 
%%% mode: latex
%%% TeX-master: "main"
%%% End: 

\section{Test run description}

\begin{table}[t]
  \centering
  \begin{tabular}{lccc}
    \hline\hline
     & TR$_0$ & TR$_1$ & TR$_2$ \\
    \hline
    input-file & sample-hmc0.input & sample-hmc2.input &
    sample-hmc3.input \\
    $L^3\times T$ & $4^3\times 4$ & $4^3\times 4$ &
    $4^3\times 4$ \\

    $S_\mathrm{G}$ & Wilson & TlSym & Iwasaki \\
    $\beta $       & $6.0$  & $3.3$ & $1.95$ \\
    $\kappa$       & $0.177$ & $0.17$ & $0.163260$ \\
    $2\kappa\mu_q$ & $0.177$ & $0.01$ & $0.002740961$ \\
    $2\kappa \bar\mu$ & $-$  & $0.1105$ & $-$ \\
    $2\kappa \bar\epsilon$ & $-$ &$0.0935$ & $-$ \\
    $\langle P\rangle$ & $0.62457(7)$ & $0.53347(17)$ & $0.5951(2)$ \\
    $\langle R\rangle$ & $-$ & $0.30393(22)$ & $0.3637(3)$ \\
    \hline\hline
  \end{tabular}
  \caption{Parameter and results for three sample input files as
    provided with the code.}
  \label{tab:testruns}
\end{table}

The source code ships with a number of sample input files. They are
located in the {\ttfamily sample-input} sub-directory. They are small
volume $V=4^4$ test runs designated to measure for instance the
average plaquette values.

Such a test-run can be performed for instance on a scalar machine by
typing 
\begin{quote}
  {\ttfamily ./hmc\_tm -f sample-hmc0.input}\ .
\end{quote}
Depending on the environment you are running in, you may need to adjust
the input parameters to match the maximal run-time and so on. The
expected average plaquette values are quoted in
table~\ref{tab:testruns} and also in the sample input files.

\subsection{Benchmark Executable}

Another useful test executable is a benchmark code. It can be build by
typing {\ttfamily make benchmark} and it will, when run, measure the
performance of the Dirac operator. It can be run in the serial or
parallel case. It reads its input from a file {\ttfamily
  benchmark.input} and the relevant input parameters are the
following:
\begin{verbatim}
L = 4
T = 4
NrXProcs = 2
NrYProcs = 2
NrZProcs = 2
UseEvenOdd = yes
UseSloppyPrecision = no
\end{verbatim}
In case of even/odd preconditioning the performance of the hopping
matrix is evaluated, in case of no even/odd the performance of the
Dirac operator. The important part of the output of the code is as
follows
\begin{verbatim}
[...]

 (1429 Mflops [64 bit arithmetic])

communication switched off
 (2592 Mflops [64 bit arithmetic])

The size of the package is 36864 Byte
The bandwidth is 662.91 + 662.91   MB/sec
\end{verbatim}
The bandwidth is not measured directly but computed from the
performance difference among with and without communication and the
package size. In case of a serial run the output is obviously
reduced.

%%% Local Variables: 
%%% mode: latex
%%% TeX-master: "main"
%%% End: 

\section{Acknowledgements}

This code was started based on a HMC implementation for Wilson
fermions with mass preconditioning kindly provided by Martin
Hasenbusch. Many discussions with, and contributions by R\'emi Baron,
Thomas Chiarappa, Albert Deuzeman, Roberto Frezzotti, Martin
Hasenbusch, Gregorio Herdoiza, Marina Marinkovic, Craig McNeile,
Istvan Montvay, Andreas Nube, David Palao, Siebren Reker, Andrea
Shindler, Jan Volkholz and Urs Wenger are gratefully acknowledged. We
thank Peter Boyle for useful discussions on the efficient
implementation of the Dirac operator..

\bibliographystyle{cpc}
\bibliography{bibliography}

\clearpage

\begin{appendix}

  \section[Gamma and Pauli matrices]{$\gamma$  and Pauli  Matrices}
  \label{sec:gammas}

In the following we specify our conventions for $\gamma$- and
Pauli-matrices. 

\subsection[Gamma-matrices]{$\gamma$-matrices}

We use the following convention for the Dirac $\gamma$-matrices:
\[
\begin{split}
  \gamma_0 = \begin{pmatrix}
    0 & 0 & -1 & 0 \\
    0 & 0 & 0 & -1 \\
    -1 & 0 & 0 & 0 \\
    0 & -1 & 0 & 0 \\
  \end{pmatrix},\quad
  \gamma_1 = \begin{pmatrix}
    0 & 0 & 0 & -i \\
    0 & 0 & -i & 0 \\
    0 & +i & 0 & 0 \\
    +i & 0 & 0 & 0 \\    
  \end{pmatrix},\\
  \gamma_2 = \begin{pmatrix}
    0 & 0 & 0 & -1 \\
    0 & 0 & +1 & 0 \\
    0 & +1 & 0 & 0 \\
    -1 & 0 & 0 & 0 \\   
  \end{pmatrix},\quad
  \gamma_3 = \begin{pmatrix}
    0 & 0 & -i & 0 \\
    0 & 0 & 0 & +i \\
    +i & 0 & 0 & 0 \\
    0 & -i & 0 & 0 \\
  \end{pmatrix}\ .\\
\end{split}
\]
In this representation $\gamma_5$ is diagonal and reads
\[
  \gamma_5 =
  \begin{pmatrix}
    +1 & 0 & 0 & 0 \\
    0 & +1 & 0 & 0 \\
    0 & 0 & -1 & 0 \\
    0 & 0 & 0 & -1 \\    
  \end{pmatrix}\ .
\]

\subsection{Pauli-matrices}

For the Pauli-matrices acting in flavour space we use the following
convention: 
\[
\begin{split}
  1_f = 
  \begin{pmatrix}
    1 & 0 \\
    0 & 1 \\
  \end{pmatrix},\quad
  \tau^1 =
  \begin{pmatrix}
    0 & 1 \\
    1 & 0 \\
  \end{pmatrix},\quad
  \tau^2 = 
  \begin{pmatrix}
    0 & -i \\
    i & 0 \\
  \end{pmatrix},\quad
  \tau^3 = 
  \begin{pmatrix}
    1 & 0 \\
    0 & -1 \\
  \end{pmatrix}
\end{split}
\]

  \section{Even/Odd Preconditioning}
  % $Id: eo-pre.tex,v 1.8 2005/09/19 11:04:51 urbach Exp $
\label{sec:eo}

\subsection{HMC Update}

In this section we describe how even/odd
\cite{DeGrand:1990dk,Jansen:1997yt} preconditioning can be used in the
HMC algorithm in 
presence of a twisted mass term. Even/odd preconditioning is
implemented in the tmLQCD package in the HMC algorithm as well as in
the inversion of the Dirac operator, and can be used optionally.

We start with the lattice fermion action in the hopping parameter
representation in the $\chi$-basis written as
\begin{equation}
  \label{eq:eo0}
    \begin{split}
    S[\chi,\bar\chi,U] = \sum_x & \Biggl\{ \bar\chi(x)[1+2i \kappa\mu\gamma_5\tau^3]\chi(x)  \Bigr. \\
    & -\kappa\bar\chi(x)\sum_{\mu = 1}^4\Bigl[ U(x,\mu)(r+\gamma_\mu)\chi(x+a\hat\mu)\bigr. \\
    & +\Bigl. \bigl. U^\dagger(x-a\hat\mu,\mu)(r-\gamma_\mu)\chi(x-a\hat\mu)\Bigr]
    \Biggr\} \\
    \equiv &\sum_{x,y}\bar\chi(x) M_{xy}\chi(y)   
  \end{split}
\end{equation}
similar to eq.~(\ref{eq:Dtm}). For convenience we define
$\tilde\mu=2\kappa\mu$. Using the matrix $M$ one can define the
hermitian (two flavour) operator:
\begin{equation}
  \label{eq:eo1}
  Q\equiv \gamma_5 M = \begin{pmatrix}
    \Qp & \\\
    & \Qm \\
  \end{pmatrix}
\end{equation}
where the sub-matrices $\Qpm$ can be factorised as follows (Schur
decomposition): 
\begin{equation}
  \label{eq:eo2}
  \begin{split}
    Q^\pm &= \gamma_5\begin{pmatrix}
      1\pm i\tilde\mu\gamma_5 & M_{eo} \\
      M_{oe}    & 1\pm i\tilde\mu\gamma_5 \\
    \end{pmatrix} =
    \gamma_5\begin{pmatrix}
      M_{ee}^\pm & M_{eo} \\
      M_{oe}    & M_{oo}^\pm \\
    \end{pmatrix} \\
    & =
    \begin{pmatrix}
      \gamma_5M_{ee}^\pm & 0 \\
      \gamma_5M_{oe}  & 1 \\
    \end{pmatrix}
    \begin{pmatrix}
      1       & (M_{ee}^\pm)^{-1}M_{eo}\\
      0       & \gamma_5(M_{oo}^\pm-M_{oe}(M_{ee}^\pm)^{-1}M_{eo})\\
    \end{pmatrix}\, .
\end{split}
\end{equation}
Note that $(M_{ee}^\pm)^{-1}$ can be
computed to be 
\begin{equation}
  \label{eq:eo3}
  (1\pm i\tilde\mu\gamma_5)^{-1} = \frac{1\mp i\tilde\mu\gamma_5}{1+\tilde\mu^2}.
\end{equation}
Using $\det(Q)=\det(\Qp)\det(\Qm)$ the following relation can be derived
\begin{equation}
  \label{eq:eo4}
  \begin{split}
    \det(\Qpm) &\propto \det(\hQpm) \\
    \hQpm &= \gamma_5(M_{oo}^\pm - M_{oe}(M_{ee}^\pm )^{-1}M_{eo})\, ,
  \end{split}
\end{equation}
where $\hQpm$ is only defined on the odd sites of the lattice. In the
HMC algorithm the determinant is stochastically estimated using pseudo
fermion fields $\phi_o$: 
\begin{equation}
  \begin{split}
    \det(\hQp \hQm) &= \int \mathcal{D}\phi_o\,\mathcal{D}\phi^\dagger_o\
    \exp(-S_\mathrm{PF})\\
    S_\mathrm{PF} &\equiv\ \phi_o^\dagger\ \left(\hQp\hQm\right)^{-1}\phi_o\, ,
  \end{split}
\end{equation}
where the fields $\phi_o$ are defined only on the odd sites of the
lattice. In order to compute the force corresponding to the effective
action $S_\mathrm{PF}$ we need the variation of $S_\mathrm{PF}$ with respect to the gauge fields
(using $\delta (A^{-1})=-A^{-1}\delta A A^{-1}$):
\begin{equation}
  \label{eq:eo5}
  \begin{split}
    \delta S_\mathrm{PF} &= -[\phi_o^\dagger (\hQp \hQm)^{-1}\delta \hQp(\hQp)^{-1}\phi_o +
    \phi_o^\dagger(\hQm)^{-1}\delta \hQm (\hQp \hQm)^{-1} \phi_o ] \\
     &= -[X_o^\dagger \delta \hQp Y_o + Y_o^\dagger \delta\hQm X_o]
  \end{split}
\end{equation}
with $X_o$ and $Y_o$ defined on the odd sides as 
\begin{equation}
  \label{eq:eo6}
  X_o = (\hQp \hQm)^{-1} \phi_o,\quad Y_o = (\hQp)^{-1}\phi_o=\hat
  \Qm X_o\ ,
\end{equation}
where $(\hQpm)^\dagger = \hat Q^\mp$ has been used. The variation of
$\hQpm$ reads
\begin{equation}
  \label{eq:eo7}
  \delta \hQpm = \gamma_5\left(-\delta M_{oe}(M_{ee}^\pm )^{-1}M_{eo} -
    M_{oe}(M_{ee}^\pm )^{-1}\delta M_{eo}\right),
\end{equation}
and one finds
\begin{equation}
  \label{eq:eo8}
  \begin{split}
    \delta S_\mathrm{PF} &= -(X^\dagger\delta \Qp Y + Y^\dagger\delta \Qm X) \\
    &= -(X^\dagger\delta \Qp Y +(X^\dagger\delta \Qp Y)^\dagger)
  \end{split}
\end{equation}
where $X$ and $Y$ are now defined over the full lattice as
\begin{equation}
  \label{eq:eo9}
  X = 
  \begin{pmatrix}
    -(M_{ee}^-)^{-1}M_{eo}X_o \\ X_o\\
  \end{pmatrix},\quad
  Y = 
  \begin{pmatrix}
    -(M_{ee}^+)^{-1}M_{eo}Y_o \\ Y_o\\
  \end{pmatrix}.
\end{equation}
In addition $\delta\Qp = \delta\Qm, M_{eo}^\dagger = \gamma_5 M_{oe}\gamma_5$ and
$M_{oe}^\dagger = \gamma_5 M_{eo}\gamma_5$ have been used. Since the bosonic part
is quadratic in the $\phi_o$ fields, the $\phi_o$ are generated at the
beginning of each molecular dynamics trajectory with
\begin{equation}
  \label{eq:eo10}
  \phi_o = \hQp r_o,
\end{equation}
where $r_o$ is a random spinor field taken from a Gaussian distribution
with norm one.

\subsubsection{Mass non-degenerate flavour doublet}

Even/odd preconditioning can also be implemented for the mass
non-degenerate flavour doublet Dirac operator $D_h$
eq.~(\ref{eq:Dh}). Denoting 
\[
Q^h = \gamma_5 D_h
\]
the even/odd decomposition is as follows
\begin{equation}
  \label{eq:Dheo}
  \begin{split}
    Q^h &=
    \begin{pmatrix}
      (\gamma_5+i\bar\mu\tau^3 -\bar\epsilon\tau^1) & Q^h_{eo}\\
      Q^h_{oe} & (\gamma_5+i\bar\mu\tau^3 -\bar\epsilon\tau^1)\\
    \end{pmatrix} \\
    &=
    \begin{pmatrix}
      Q^h_{ee} & 0 \\
      Q^h_{oe} & 0 \\
    \end{pmatrix}
    \cdot
    \begin{pmatrix}
      1 & (Q^h_{ee})^{-1}Q_{eo} \\
      0 & Q^h_{oo} \\
    \end{pmatrix} \\
  \end{split}
\end{equation}
where $Q^h_{oo}$ is given in flavour space by
\begin{equation*}
  Q^h_{oo} = \gamma_5
  \begin{pmatrix}
    1 + i\bar\mu\gamma_5 -
    \frac{M_{oe}(1-i\bar\mu\gamma_5)M_{eo}}{1+\bar\mu^2-\bar\epsilon^2} & 
    -\bar\epsilon\left(1+\frac{M_{oe}M_{eo}}{1+\bar\mu^2-\bar\epsilon^2}\right) \\
    -\bar\epsilon\left(1+\frac{M_{oe}M_{eo}}{1+\bar\mu^2-\bar\epsilon^2}\right) & 
    1 - i\bar\mu\gamma_5 -
    \frac{M_{oe}(1-i\bar\mu\gamma_5)M_{eo}}{1+\bar\mu^2-\bar\epsilon^2}\\
  \end{pmatrix}
\end{equation*}
with the previous definitions of $M_{eo}$ etc. The implementation for
the PHMC is very similar to the mass degenerate HMC case.

\subsection{Inversion}

In addition to even/odd preconditioning in the HMC algorithm as
described above, it can also be used to speed up the inversion of the
fermion matrix. Due to the factorisation (\ref{eq:eo2}) the full
fermion matrix can be inverted by inverting the two matrices appearing
in the factorisation
\[
\begin{pmatrix}
  M_{ee}^\pm & M_{eo} \\
  M_{oe}    & M_{oo}^\pm \\
\end{pmatrix}^{-1}
=
\begin{pmatrix}
  1       & (M_{ee}^\pm)^{-1}M_{eo}\\
  0       & (M_{oo}^\pm-M_{oe}(M_{ee}^\pm)^{-1}M_{eo})\\
\end{pmatrix}^{-1}
\begin{pmatrix}
  M_{ee}^\pm & 0 \\
  M_{oe}   & 1 \\
\end{pmatrix}^{-1}\, .
\]
The two factors can be simplified as follows:
\[
\begin{pmatrix}
  M_{ee}^\pm & 0 \\
  M_{oe}   & 1 \\
\end{pmatrix}^{-1}
=
\begin{pmatrix}
      (M_{ee}^\pm)^{-1} & 0 \\
      -M_{oe} (M_{ee}^{\pm})^{-1}  & 1 \\
    \end{pmatrix}
\]
and 
\[
\begin{split}
  &\begin{pmatrix}
    1       & (M_{ee}^\pm)^{-1}M_{eo}\\
    0       & (M_{oo}^\pm-M_{oe}(M_{ee}^\pm)^{-1}M_{eo})\\
  \end{pmatrix}^{-1}
  \\=&
  \begin{pmatrix}
    1       & -(M_{ee}^\pm)^{-1}M_{eo}(M_{oo}^\pm-M_{oe}(M_{ee}^\pm)^{-1}M_{eo})^{-1}  \\
    0       & (M_{oo}^\pm-M_{oe}(M_{ee}^\pm)^{-1}M_{eo})^{-1}\\
  \end{pmatrix}\, .
\end{split}
\]
The complete inversion is now performed in two separate steps: first
compute for a given source field $\phi=(\phi_e,\phi_o)$ an intermediate 
result $\varphi=(\varphi_e,\varphi_o)$ by:
\[
\begin{pmatrix}
  \varphi_e \\ \varphi_o\\
\end{pmatrix}
=
\begin{pmatrix}
  M_{ee}^\pm & 0 \\
  M_{oe}   & 1 \\
\end{pmatrix}^{-1}
\begin{pmatrix}
  \phi_e \\ \phi_o \\
\end{pmatrix}
=
\begin{pmatrix}
  (M_{ee}^\pm)^{-1} \phi_e \\ 
  -M_{oe}( M_{ee}^\pm)^{-1} \phi_e + \phi_o \\
\end{pmatrix}\, .
\]
This step requires only the application of $M_{oe}$ and
$(M_{ee}^\pm)^{-1}$, the latter of which is given by eq.~(\ref{eq:eo3}).
The final solution $\psi=(\psi_e,\psi_o)$ can then be computed with
\[
\begin{pmatrix}
  \psi_e \\ \psi_o \\
\end{pmatrix}
=
\begin{pmatrix}
  1       & (M_{ee}^\pm)^{-1}M_{eo}\\
  0       & (M_{oo}^\pm-M_{oe}(M_{ee}^\pm)^{-1}M_{eo})\\
\end{pmatrix}^{-1}
\begin{pmatrix}
  \varphi_e \\ \varphi_o \\
\end{pmatrix}
=
\begin{pmatrix}
  \varphi_e - (M_{ee}^\pm)^{-1}M_{eo}\psi_o \\ \psi_o \\
\end{pmatrix}\, ,
\]
where we defined
\[
\psi_o = (M_{oo}^\pm-M_{oe}(M_{ee}^\pm)^{-1}M_{eo})^{-1} \varphi_o\, .
\]
Therefore, the only inversion that has to be performed numerically is
the one to generate $\psi_o$ from $\varphi_o$ and this inversion
involves only an operator that is better conditioned than the original
fermion operator.

Even/odd preconditioning can also be used for the mass non-degenerate
Dirac operator $D_h$ eq.~(\ref{eq:Dh}). The corresponding equations
follow immediately from the previous discussion and the definition
from eq.~(\ref{eq:Dheo}).

  \section{Initialising the PHMC}
  \label{sec:root}

The function $1/\sqrt{s}$ in the interval $[\epsilon,1]$ can be
approximated using polynomials or rational functions of different
sorts. In the tmLQCD package we use Chebysheff polynomials, which
are easy to construct. They can be constructed as to provide a desired
overall precision in the interval $[\epsilon,1]$. 

As discussed in sub-section~\ref{sec:ndphmc}, the roots of the
polynomial $P_{n,\epsilon}$ are needed for the evaluation of the
force. Even though the roots come in complex conjugate pairs, for our
case the roots cannot be computed analytically, hence we need to
determine them numerically. Such an evaluation requires usually
high precision. This is why these roots need to be determined
\emph{before} a PHMC run using an external program, i.e. they cannot
be computed at the beginning of a run in the {\ttfamily hmc\_tm}
program. 

Such an external program ships with the tmLQCD code, which is located in the
{\ttfamily util/laguere} directory\footnote{We thank Istvan Montvay
  for providing us with his code.}. It is based on Laguerre's
method and uses the Class Library for Numbers
(CLN)~\cite{cln:web}, which provides arbitrary precision data
types. In order to compute roots the CLN library must be available,
which is free software.

Taking for granted that the CLN library is available,
the procedure for computing the roots is as follows: assuming the
non-degenerate Dirac operator has eigenvalues in the interval $[\tilde
s_\mathrm{min},\tilde s_\mathrm{max}]$, i.e. $\epsilon = \tilde
s_\mathrm{min}/\tilde s_\mathrm{max}$, and the polynomial degree is
$n$. Edit the file {\ttfamily chebyRoot.H} and set the
variable {\ttfamily EPSILON} to the value of $\epsilon$. Moreover,
set the variable {\ttfamily MAXPOW} to the degree $n$. Adapt the
{\ttfamily Makefile} to your local installation and compile the code
by typing {\ttfamily make}. After running the {\ttfamily ChebyRoot}
program successfully, you should find two files in the directory
\begin{enumerate}
\item {\ttfamily Square\_root\_BR\_roots.dat}:\\
  which contains the roots of the polynomial in bit-reverse
  order~\cite{Frezzotti:1997ym}.
\item {\ttfamily normierungLocal.dat}:\\
  which contains a normalisation constant. 
\end{enumerate}
Copy these two files into the directory where you run the code and
adjust the input parameters to match \emph{exactly} the values used
for the root computation. I.e. the input parameters {\ttfamily
  StildeMin}, {\ttfamily StildeMax} and {\ttfamily
  DegreeOfMDPolynomial} must be set appropriately in the {\ttfamily
  NDPOLY} monomial.

The minimal and maximal eigenvalue of the non-degenerate flavour
doublet can be computed as an online measurement. The frequency can be
specified in the {\ttfamily NDPOLY} monomial with the input parameter
{\ttfamily ComputeEVFreq} and they are written to the file called
{\ttfamily phmc.data}. Note that this is not a cheap operation in
terms of computer time. However, if the approximation interval of the
polynomial is chosen wrongly the algorithm performance might
deteriorate drastically, in particular if the upper bound is set
wrongly. It is therefore advisable to introduce some security measure
in particular in the value of $\tilde s_\mathrm{max}$. 

%%% Local Variables: 
%%% mode: latex
%%% TeX-master: "main"
%%% End: 

\end{appendix}

\end{document}